\documentclass[a4paper,fleqn,usenatbib]{mnras}

\usepackage{amsmath}	
\usepackage{amssymb}	
\usepackage{txfonts}

\usepackage[T1]{fontenc}
\usepackage{ae,aecompl}

\usepackage{graphicx}	
\usepackage{natbib}
\usepackage{url}

\usepackage{multirow}

\bibpunct[; ]{(}{)}{;}{a}{}{,}

\newcommand{\eqnref}[1]{equation~(\ref{#1})}

\newcommand{\Herschel}{\textit{Herschel}}
\newcommand{\tphot}{\textsc{t-phot}}

\newcommand{\Lir}{\ensuremath{L_\text{IR}}}
\newcommand{\Luv}{\ensuremath{L_\text{UV}}}
\newcommand{\Lsun}{\ensuremath{\textrm{L}_{\sun}}}
\newcommand{\sfr}{\ensuremath{\textrm{M}_{\sun} \textrm{yr}^{-1}}}
\newcommand{\Mstar}{\ensuremath{M_{\ast}}}
\newcommand{\Msun}{\ensuremath{\textrm{M}_{\sun}}}
\newcommand{\MUV}{\ensuremath{M_\text{UV}}}

\title[Evolution of cosmic star formation]{Evolution of cosmic star formation in the SCUBA-2 Cosmology Legacy Survey}
\author[N. Bourne et al.]
 {N. Bourne,$^{1}$\thanks{nbourne22@gmail.com}
 J.\,S. Dunlop,$^{1}$
 E.~Merlin,$^2$
 S.~Parsa,$^1$
 C.~Schreiber,$^3$
 M.~Castellano,$^2$\newauthor
 C.\,J.~Conselice,$^{4}$
 K.\,E.\,K.~Coppin,$^{5}$
 D.~Farrah,$^6$
 A.~Fontana,$^2$
 J.\,E.~Geach,$^{5}$\newauthor
 M.~Halpern,$^7$
 K.\,K.~Knudsen,$^8$
 M.\,J.~Micha{\l}owski,$^1$
 A.~Mortlock,$^1$
 P.~Santini,$^2$\newauthor
 D.~Scott,$^7$
 X.\,W.~Shu,$^{9}$
 C.~Simpson,$^{10}$
 J.\,M.~Simpson,$^1$
 D.\,J.\,B.~Smith,$^{5}$\newauthor
 P.\,P.~van~der~Werf\,$^{3}$
   \\
   $^1$Institute for Astronomy, University of Edinburgh, Royal Observatory, Edinburgh EH9 3HJ, UK\\
   $^2$INAF-Osservatorio Astronomico di Roma, via di Frascati 33, I-00078 Monte Porzio Catone, Roma, Italy\\
   $^3$Leiden Observatory, Leiden University, PO Box 9513, 2300 RA Leiden, The Netherlands\\
   $^{4}$School of Physics and Astronomy, University of Nottingham, University Park, Nottingham NG7 2RD, UK\\
   $^{5}$Centre for Astrophysics Research, Science and Technology Research Institute, University of Hertfordshire, Hatfield AL10 9AB, UK\\
   $^6$Department of Physics, Virginia Tech, Blacksburg, VA 24061, USA\\
   $^7$Department of Physics and Astronomy, University of British Columbia, Vancouver, Canada\\
   $^8$Department of Earth and Space Sciences, Chalmers University of Technology, Onsala Space Observatory, SE-43992 Onsala, Sweden\\
   $^9$Department of Physics, Anhui Normal University, Wuhu, Anhui, 241000, China\\
   $^{10}$Gemini Observatory, Northern Operations Center, 670 N.~A`oh\={o}k\={u} Place, Hilo, HI 96720-2700, USA
 }

\date{Accepted XXX. Received YYY; in original form ZZZ}

\pubyear{2016}

\begin{document}
\label{firstpage}
\pagerange{\pageref{firstpage}--\pageref{lastpage}}
\maketitle

\begin{abstract}
We present a new exploration of the cosmic star-formation history and dust obscuration in massive galaxies at redshifts $0.5<z<6$. We utilize the deepest 450 and 850\micron\ imaging from SCUBA-2 CLS, covering 230~arcmin$^2$ in the AEGIS, COSMOS and UDS fields, together with 100--250\micron\ imaging from \Herschel. 
We demonstrate the capability of the \tphot\ deconfusion code to reach below the confusion limit, using multi-wavelength prior catalogues from CANDELS/3D-HST. By combining IR and UV data, we measure the relationship between total star-formation rate (SFR) and stellar mass up to $z\sim5$, indicating that UV-derived dust corrections underestimate the SFR in massive galaxies. 
We investigate the relationship between obscuration and the UV slope (the IRX--$\beta$ relation) in our sample, which is similar to that of low-redshift starburst galaxies, although it deviates at high stellar masses. 
Our data provide new measurements of the total SFR density (SFRD) in $\Mstar>10^{10}\Msun$ galaxies at $0.5<z<6$. This is dominated by obscured star formation by a factor of $>10$. One third of this is accounted for by 450\micron-detected sources, while one fifth is attributed to UV-luminous sources (brighter than $\Luv^\ast$), although even these are largely obscured. 
By extrapolating our results to include all stellar masses, we estimate a total SFRD that is in good agreement with previous results from IR and UV data at $z\lesssim3$, and from UV-only data at $z\sim5$.
The cosmic star-formation history undergoes a transition at $z\sim3-4$, as predominantly unobscured growth in the early Universe is overtaken by obscured star formation, driven by the build-up of the most massive galaxies during the peak of cosmic assembly.
\end{abstract}




\section{Introduction}

\label{sec:intro}
A key element in understanding the evolution of galaxies and the build-up of the present-day population is the cosmic star-formation history, i.e. the overall comoving volume-density of the star-formation rate (SFR) within galaxies throughout the Universe, measured as a function of look-back time. This has been observationally determined from ultraviolet (UV) emission from star-forming galaxies up to $z\sim9$, within the first few hundred Myr of the Universe
\citep{Bouwens2007,Reddy2008,Cucciati2012,McLure2013,Duncan2014,Bouwens2015,Mashian2015,McLeod2015,McLeod2016,Parsa2015}.  
{It is well known, however, that over most of cosmic history} the great majority of the UV radiation from young stars is absorbed by dust within galaxies and is thermally re-radiated in the far-infrared (FIR; \citealp{Desert1990}). 
{As a result, rest-frame UV observations must either be corrected for these extinction losses, or supplemented with observations in the rest-frame FIR, to recover the total SFR.}
{The cosmic star-formation history can also be explored through measurements of the average specific SFR (SSFR~=~SFR/stellar mass) of galaxy samples. This is known to increase with look-back time, but may plateau or rise more slowly at $z>3$ \citep{Madau2014}. 
Meanwhile, stellar mass density is constrained out to high redshifts with smaller systematic errors due to the reduced dust extinction effects in the rest-frame near-IR \citep{Ilbert2013, Muzzin2013,Grazian2015}.
This must approximately trace the total SFR integrated over all masses and times up to a given look-back time, so we can expect to see the rate of growth of stellar mass trace the cosmic SFR evolution \citep{Wilkins2008, Madau2014}.}

{While the large body of observational work appears to have converged on a consistent picture of the cosmic star-formation history \citep{Behroozi2013b,Madau2014}, there remain
significant discrepancies with the most up-to-date hydrodynamical simulations and semi-analytic models. Both flavours of physical models are unable to consistently explain the evolution of the cosmic SFR density (SFRD), the history of stellar-mass assembly, and the average SFRs of individual galaxies as a function of their stellar mass; suggesting that either there must be systematic errors in the observational results, or the models do not adequately describe properties such as the dust attenuation curve 
\citep{Somerville2012,Kobayashi2013,Genel2014,Furlong2015, Lacey2016}. 
}

{The obscuration of star formation by dust has been studied extensively using \textit{Herschel} and \textit{Spitzer} FIR data \citep{Buat2012, Reddy2012a, Burgarella2013a, Heinis2014, Price2014}. It has been shown to correlate with stellar mass and SFR in galaxy samples spanning a wide range of redshifts \citep{Reddy2010,Garn2010,Hilton2012a,Heinis2014,Price2014,Whitaker2014,Pannella2015}.
However, these scaling relations contain significant scatter and we must seek more physically meaningful observations to allow us to predict the dust obscuration in the absence of direct FIR measurements.
The empirical relationship which is usually employed for this task is the so-called ``IRX--$\beta$'' relation, which connects the obscuration fraction (FIR/UV luminosity ratio; IRX) and the observed slope ($\beta$) of the UV spectral energy distribution (SED). 
The connection is motivated by the principle that the same dust is responsible for the reddening of the intrinsic SED and for the reprocessing of the extincted UV flux into the FIR. 
The exact relationship is calibrated empirically on low-redshift samples \citep{Meurer1999, Calzetti2000, Kong2004, Boissier2007, Munoz-Mateos2009, Overzier2011, Hao2011, Takeuchi2012}, but is by no means fundamental. 
It depends on the UV attenuation curve and the intrinsic UV spectral slope (hence may be sensitive to metallicity, star-formation history, and the initial mass function), and it relies on the basic assumption of a simple dust screen obscuring all of the UV emission isotropically. 

Low-redshift calibrations of the IRX--$\beta$ relation \citep[such as the starburst calibration of][]{Meurer1999} are
important for the study of star formation at high redshifts, where rest-frame UV data are frequently the only available tracer in large samples; yet it remains unclear whether these assumptions are valid across the full star-forming galaxy population at high redshifts
\citep{Gonzalez-Perez2013, Castellano2014, Heinis2014, Price2014,Yajima2014, Capak2015, Coppin2015, Pannella2015, Zeimann2015, Talia2015, Watson2015}.
}
The chief problem {to overcome} is the difficulty of measuring accurate rest-frame FIR emission in representative samples of star-forming galaxies at $z>3$, as a result of the high confusion noise in sub-millimetre (submm) surveys with moderate-sized, single-dish telescopes such as \Herschel\
\citep{Reddy2012a,Burgarella2013a,Gruppioni2013,Heinis2014}.
Interferometric imaging surveys of deep fields (e.g. with ALMA) offer a high-resolution alternative to single-dish surveys \citep{Bouwens2016, Kohno2016, Hatsukade2016, Dunlop2016}, but these only probe relatively small volumes which have limited statistical power.

Apart from biased samples of bright Lyman-alpha emitters and submm-selected starbursts (which are rare and therefore unrepresentative of the overall star-forming-galaxy population), the only significant source of information on the cosmic SFR at early times comes from samples of photometrically selected Lyman-break galaxies (LBGs). 
These samples can be used to measure the rest-frame UV luminosity function (LF), from which the total {SFRD} can be extrapolated by integrating the LF model and applying (potentially large) corrections for dust obscuration.
The obscuration of star formation in $z\gtrsim3$ LBGs has been studied in the submm via stacking of SCUBA-2 and \Herschel\ data by \citet{Coppin2015}; 
{AzTEC and \Herschel\ data by \citet{Alvarez-Marquez2016};}
and individually using ALMA and the Plateau de Bure Interferometer \citep[e.g.][and see also \citealp{Chapman2000, Peacock2000, Stanway2010, Davies2012, Davies2013}]{Schaerer2015,Capak2015,Bouwens2016}, but the question of whether low-redshift calibrations hold true {for all star-forming galaxies at high redshifts remains open \citep[see for example the recent discussion by][]{Alvarez-Marquez2016}.}

It is clear {from our knowledge of the UV and FIR luminosity densities \citep[e.g.][]{Burgarella2013a,Madau2014}} that the vast majority of the SFR in the Universe is obscured by dust, and this fraction appears to increase from $z=0$ to $z=1$. However, the behaviour at higher redshifts is uncertain \citep{Burgarella2013a}.
There are systematic uncertainties in the true behaviour of the dust-obscured (hence total) SFRD at high redshifts due to uncertainties about the nature of star-forming galaxies at high redshifts.
Typical star-forming galaxies at high redshifts have higher SSFRs than their counterparts at low redshifts, so it is unclear whether they resemble their low-redshift counterparts (which might be implied by the existence of a common mass-SFR relation known as the ``main sequence''; \citealt{Noeske2007}) or whether they are more similar to high-SSFR galaxies at low redshifts (because they are similarly rich in dense gas and have a clumpy mass distribution; \citealt{Price2014}). 

In summary we need to improve our knowledge of the obscured SFRD from FIR observations at $z>3$. Currently we are limited by issues including sample bias 
{(towards unrepresentative bright objects selected in the FIR, or more unobscured objects selected in the UV), and uncertainties due to scatter within the population when stacking fainter objects (such as those selected by stellar mass).}
In the FIR, the major obstacle is the low resolution of single-dish FIR/submm surveys (typically 15--35 arcsec), which limits our ability to detect and identify individual sources above the confusion limit, and also hampers stacking below the confusion limit due to the difficulty of separating the emission from heavily blended positions.

In this paper we attempt to improve this situation with a combination of three key ingredients: (i) deep, high resolution submm imaging of blank fields with JCMT/SCUBA-2 \citep{Holland2013}; (ii) rich multi-wavelength catalogues containing positions, redshift information and UV-to-mid-IR SEDs to support the submm data; (iii) the latest deconfusion techniques developed within the ASTRODEEP consortium, which allow us to maximize the useful information available from these combined data sets.
In Section~\ref{sec:data} we briefly describe the submm imaging data and the sample used in this work. In Section~\ref{sec:methods} we explain how \tphot\ is applied to measure deconfused submm photometry for the sample, and in Section~\ref{sec:analysis} we describe and validate the stacking technique. Section~\ref{sec:results} presents the results and discussion in the context of previous literature.
Throughout this paper we assume a flat $\Lambda$CDM cosmology with $\Omega_M=0.3$, $h=H_0/100$~km\,s$^{-1}$\,Mpc$^{-1}=0.7$. All magnitudes are in the AB system \citep{Oke1974,Oke1983} and we assume the \citet{Kroupa2003} initial mass function (IMF) throughout, unless otherwise stated.

\section{Data}
\label{sec:data}
We use data from the SCUBA-2 Cosmology Legacy Survey \citep[S2CLS;][]{Geach2013,Geach2016,Roseboom2013}, which provides imaging over $\sim5$ deg$^2$ at 850~\micron\ in several wide/deep survey fields, as well as $\sim0.25$ deg$^2$ at 450 and 850~\micron\ in several ultra-deep fields, which benefit from multi-wavelength coverage from CANDELS \citep{Grogin2011,Koekemoer2011}.

In order to probe the SFRs of the faint and highly-confused source population at the highest redshifts we require the best possible submm resolution (from a single-dish blind survey) to minimise confusion noise, and the deepest available imaging to minimise instrumental noise.
This is provided by the 450-\micron\ imaging of the CANDELS COSMOS, UDS and AEGIS fields, covering a total of 230 arcmin$^2$ to depths of $\sim1.0$~mJy\,beam$^{-1}$ rms (average instrumental noise in match-filtered maps) with resolution FWHM~$=7.5$~arcsec.
Due to its high angular resolution and long wavelength, the 450-\micron\ imaging is central to our analysis.
However, we additionally benefit from 
SCUBA-2 850-\micron\ imaging with rms~$\sim0.2$mJy\,beam$^{-1}$ (match-filtered), FWHM~$=14$~arcsec;
\Herschel/PACS imaging with rms~$\sim1$--$2$~mJy\,beam$^{-1}$
and FWHM~$=9,11$~arcsec at 100 and 160~\micron\ respectively (from PEP and HerMES; \citealp{Lutz2011,Oliver2012});
and \Herschel/SPIRE imaging with rms~$\sim2$--$4$~mJy\,beam$^{-1}$ 
and FWHM~$=18$~arcsec at 250~\micron\ (from HerMES).
The SCUBA-2 maps have been reduced using the \textsc{smurf} pipeline described by \citet{Chapin2013}, which includes a band-pass filter in the time series equivalent to angular scales of 2 to 120 arcsec \citep[for further details see][]{Geach2013,Geach2016}. However, the maps that we use have not been processed with a matched filter since our methodology effectively performs this filtering independently. The flux-conversion factors applied to the 450 and 850-\micron\ data are 540 and 591~Jy\,beam$^{-1}$\,pW$^{-1}$; these are the canonical values with an additional 10 per cent correction to account for losses in the filtering procedure \citep{Chapin2013, Dempsey2013}. 

We use multi-wavelength catalogues compiled by the 3D-HST team \citep{Brammer2012, Skelton2014}, which include photometry spanning the $u$--8\micron\ bands from CFHTLS, Subaru, CANDELS, NMBS, WIRDS, UKIDSS-UDS, UltraVISTA, SEDS, S-COSMOS and EGS. 
These catalogues are derived from {\it HST}/WFC3 imaging and are effectively $H$-band selected from imaging with a median \mbox{$5\sigma$} depth of $H({\rm F160W})=26.4$ in a 1-arcsec aperture; the catalogues have 50 and 95 per cent completeness limits of $H=26.5$ and $25.1$ respectively \citep{Skelton2014}.
We use the combined ``Z\_BEST'' redshift data from 3D-HST \citep{Momcheva2015} which comprise collected spectroscopic redshifts from the literature, {\it HST} grism redshifts where reliable, and photometric redshifts from \textsc{eazy} \citep{Brammer2008} otherwise. 
The spectroscopic and photometric redshifts are described by \citet{Skelton2014}; \textsc{eazy} photometric redshifts are based on SED fitting with a linear combination of seven templates, which consist of the five default \textsc{p{\'e}gase} stellar population synthesis (SPS) models from \citet{Fioc1997}, in addition to a young, dusty template and an old, red template from \citet{Whitaker2011}.
Grism redshifts are described by \citet{Momcheva2015}, and are based on observations with the WFC3 G141 grism covering 60 per cent of the CANDELS imaging in AEGIS, COSMOS and UDS, for photometric objects selected down to a depth of $JH=26$ in the co-added F125W+F140W+F160W images. Redshifts were determined from fitting the 2D spectra and multi-band photometry simultaneously, using a modified version of the \textsc{eazy} templates with additional emission-line templates from \citet{Dobos2012}. 
Redshift fits to all grism spectra were visually inspected for quality, and typical redshift uncertainties are $\Delta z/(1+z)\approx 0.003$ \citep{Momcheva2015}. Photometric redshift uncertainties are typically $\Delta z/(1+z)\leq 0.03$ (and $\leq0.01$ in COSMOS due to good medium-band coverage), with fewer than 5 per cent significant outliers in all fields \citep{Skelton2014}. Our final sample (see Section~\ref{sec:selection}) consists of 6 per cent spectroscopic redshifts, 49 per cent grism redshifts and 45 per cent photometric redshifts. Most of the redshifts at $z>3$ are photometric, but this subset also has typical uncertainties $\leq0.03$.

\citet{Skelton2014} also provide stellar-population parameters derived from SED fitting with \textsc{fast} \citep{Kriek2009}.
These are based on \citet{Bruzual2003} SPS models with a \citet{Chabrier2003} IMF;\footnote{We make no adjustment for the difference between \citet{Chabrier2003} and \citet{Kroupa2003} IMFs, which are essentially identical for our purposes \citep{Chomiuk2011}.} 
solar metallicity; exponentially-declining star-formation histories with a minimum e-folding time of $10^7$~yr; minimum age of 40~Myr; $0<A_V<4$ and a \citet{Calzetti2000} dust attenuation law.
We use the stellar masses from these SED fits (which are well-constrained by the available optical-near-IR photometry; \citealp{Skelton2014}), and we use $\MUV=M_{1600\AA}$ (i.e. FUV-band) estimated from the rest-frame SED fit from \textsc{eazy} \citep{Skelton2014}. At this wavelength, $\MUV$ traces emission from stellar populations with a mean age of $10^7$yr \citep{Kennicutt2012}.
More detailed sample selection and binning is described in Section~\ref{sec:selection}.

\section{Methods}
\label{sec:methods}

Many different algorithms have been developed to solve the problem of deconfusing low-resolution imaging using prior information from high-resolution surveys \citep{Bethermin2010, Kurczynski2010a, Roseboom2010, Bourne2012, Viero2013, Wang2014, Mackenzie2015, Safarzadeh2015, Hurley2016, Wright2016}.
In this study we use \tphot\ \citep{Merlin2015,Merlin2016}\footnote{\tphot\ is a multi-purpose deconfusion code applicable to multi-wavelength data sets with a range of angular resolutions; it was developed within the ASTRODEEP consortium and is available from \url{http://astrodeep.eu}}
to fit submm low-resolution images (LRI) with positional priors from higher-resolution data-sets. 
\tphot\ is a versatile tool which can be applied to a wide range of problems as described by \citet{Merlin2015}. In general it provides a fast and efficient algorithm for measuring deblended photometry in low-resolution imaging based on prior information from higher-resolution data-sets, which may consist of high-resolution images, parametric light-profile models, and/or simply positional priors, depending on the particular data-set in question.
The specific application of \tphot\ to confusion-limited, unresolved submm images such as those from SCUBA-2 involves a number of elements which differ significantly from applications in less confused optical-to-mid-IR images.
These can be broken down into the following four-stage procedure:
\begin{enumerate}
\item {\bf Selection:} select a suitable source catalogue to use for the positional priors.
\item {\bf Input:} provide as inputs the list of prior positions, the LRI and associated rms noise map, and the PSF.
\item {\bf Optimization:} construct a model of the entire LRI consisting of scaled point sources at every prior position, and obtain a set of flux measurements (i.e. the best-fitting normalisation for each point source) that minimizes $\chi^2$ between this model and the LRI.
\item {\bf Background:} measure the overall background level of the LRI, and account for this in the flux measurements of sources. The latest version of \tphot\ \citep{Merlin2016} fits the background as a free parameter in the model. We investigate the accuracy of the background measurement and whether any further refinement is required.
\end{enumerate}
To fully explain the application of \tphot\ to confused submm images with dense prior lists, we expand on of each of these steps in the following subsections (\ref{sec:selection}--\ref{sec:bgs}).

\subsection{Selection}
\label{sec:selection}
The definition of the prior catalogue is of great importance when attempting to model a confused map by $\chi^2$ minimisation. 
However, obtaining a formally good fit does not guarantee accurate estimates for the fluxes or their associated errors or covariances: this relies on an appropriate set of priors being used. For example, if an object which is bright in the LRI is not represented in the priors then the fluxes of any nearby priors will be over-estimated due to blending with the additional bright object not being accounted for, and this systematic error will not be included in the covariance matrix \citep[see][]{Merlin2015}.

We select priors from the 3D-HST parent catalogue (which is effectively $H$-band selected -- see Section~\ref{sec:data}) by first imposing limits in AB magnitude of $K_s<24$ or IRAC [3.6]$<24$,  which are chosen to maximise the completeness of massive galaxies at $z<6$ (we analyse the completeness in Section~\ref{sec:comp}).  
{The combined $K_s$ and [3.6] photometric selection ensures that the rest-frame selection wavelength is $>5000\AA$ over the full redshift range, which is primarily sensitive to stellar mass, and relatively insensitive to variables such as star-formation history or dust obscuration.
}
We include only galaxies with the ``USE'' photometric flag from 3D-HST,
which ensures sufficient photometry for a photometric redshift fit.
The USE flag excludes stars; objects whose photometry may be affected by nearby bright stars; objects which don't have at least two individual exposures in F125W or F160W; objects with F160W signal-to-noise ratio $\leq3$; objects with catastrophic photometric-redshift fits ($\chi^2\geq1000$) or catastrophic stellar-population fits ($\log \Mstar <0$). The fraction of objects in the 3D-HST catalogues of AEGIS, COSMOS and UDS which have the USE flag is 87 per cent, and this fraction is essentially independent of magnitude at F160W$<26.5$.

To avoid excessive crowding of the priors at lower redshifts we also impose a cut on stellar mass of $\log{(\Mstar/\Msun)}>9$. In this way we deblend only massive galaxies in the model; galaxies of lower mass become part of the background to be subtracted (see Section~\ref{sec:bgs}). 
By restricting the priors in this way, we implicitly assume that any population of objects that contributes to the map, but is missing from the priors, is not spatially correlated with the priors. Clearly this assumption is flawed when considering that low-mass galaxies will be clustered around high-mass galaxies, but we can justify this decision as follows.
The contribution of $\log{(\Mstar/\Msun)}<9$ galaxies to the 450-\micron\ sky (i.e. the cosmic infrared background, CIB) will be relatively small compared to higher-mass galaxies, assuming (i) that 450-\micron\ flux is roughly proportional to SFR at $z<6$ \citep{Blain2002},\footnote{In fact there is a redshift-dependence, but it is weak at \mbox{$1<z<4$}, and the only lower-mass galaxies with significant 450-\micron\ emission would have to be either at low redshift, or have extremely high SFR/$\Mstar$ ratios.}
 (ii) that SFR, in particular obscured SFR, is roughly proportional to stellar mass \citep[e.g.][]{Noeske2007,Elbaz2011}, and (iii) that most of the stellar mass resides in galaxies more massive than $10^9\Msun$ \citep{Kauffmann2003, Kajisawa2009, Marchesini2009, Mortlock2011}.
The stellar mass threshold ensures that any bias incurred from this assumption will be roughly invariant with redshift, whereas if we used only a magnitude cut the bias would increase as the mass cut rises with increasing redshift.
In order to test for potential bias as a result of the mass limit, we tested the effect of using a higher limit of $\log{(\Mstar/\Msun)}=9.5$. If there is a significant bias due to flux from lower-mass galaxies (outside of the sample) being attributed to galaxies in the sample, that bias would be larger when we increase the log mass limit from 9.0 to 9.5. In fact, we find that the results obtained with the two different mass limits are fully consistent with each other, and all of the measured trends that we discuss are robust.

Our final sample consists of 8809 positional priors within an area of 230~arcmin$^2$ over the AEGIS, COSMOS and UDS fields. 

\subsection{Input}
The inputs for \tphot\ are specified within the parameter file as described by \citet{Merlin2015}. In this case we provide the 450-\micron\ image of each field (the LRI), along with the rms instrumental noise map from the SCUBA-2 pipeline \citep{Chapin2013}, and the PSF. The data have been band-pass-filtered as part of the map-making procedure in order to remove large-scale background variations, and the maps have zero mean \citep{Geach2013, Chapin2013}. We use images that are not PSF-filtered or match-filtered because the \tphot\ algorithm itself effectively filters the image and deconfuses all sources, so there is no benefit to using additional filtering. 
The PSF of these pre-matched-filter images is therefore assumed to be a symmetrical Gaussian function with FWHM=7.5~arcsec \citep{Chapin2013}.
The prior catalogue is also provided as a list of $x,y$ positions relative to the image.

The 450-\micron\ data provide the backbone of our analysis, but in order to characterise FIR/submm SEDs we
also apply \tphot\ to images at 100, 160, 250 and 850\micron. We assume {PSFs} with FWHM of 9, 11, 18 and 14 arcsec in each of these bands respectively. 
The larger beam sizes in these lower-resolution images means that confusion-related uncertainties are much larger, as a result of the far greater degeneracies between highly blended galaxies. However, the uncertainties output by \tphot\ account for the covariances between blended galaxies, and therefore the output measurements can be successfully combined to constrain the average properties of samples at these wavelengths, as described in Section~\ref{sec:sedfit}.

When fitting confused maps in this way, it is of vital importance that the map is appropriately masked, since any sources lying outside the region covered by the priors will not be included in the model and can easily dominate the residual, leading to degenerate $\chi^2$ values. We therefore ensure that the prior catalogues and images are matched, 
by masking the outer parts of the image that are at least 10 pixels (approximately $3\times$FWHM of the PSF) outside the area covered by the prior catalogue.

\subsection{Optimization}
We run \tphot\ using the recommended options for unresolved priors: we use a single pass, since the precise astrometric positions of the priors cannot be improved by allowing them to shift in the ``dance'' stage;\footnote{The ``dance'' stage allows for precise re-registration of image priors to account for astrometric shifts between bands, and would involve a second pass of the optimization routine to improve the photometry \citep{Merlin2015}.} we fit the entire image with a single model rather than dividing it into cells; and we allow \tphot\ to fit a single-valued sky background as a free parameter (see Section~\ref{sec:bgs}). 
An example of a \tphot\ parameter file used in this analysis is provided in Appendix~\ref{app:tphotparfile}.
Once the best-fitting model has been obtained, \tphot\ outputs a model image comprising a ``collage'' of all the priors with their best-fitting normalizations (flux densities), a residual given by the difference LRI~--~model, a catalogue of best-fitting parameters for each object, and the covariance matrix of the best fit. The fluxes in the catalogue are background-subtracted, while the model and residual can be background-subtracted using the output background value in the \tphot\ log file.

\subsection{Background}
\label{sec:bgs}
The sky background (i.e. any signal that remains in the image after the individual sources under consideration have been subtracted) is an important property of the image, especially when we want to measure very faint sources close to and below the confusion noise level. The current version of \tphot\ includes the background as a free parameter in the model, and in this section we explore the accuracy and uncertainty of this fit.

Background subtraction in general is a problem for confusion-limited maps, which have no empty regions of sky in which to measure the background. For this reason, such maps are usually set to have zero mean, such that positive bright sources and over-dense regions are balanced by negative surface brightness in regions where the source density is low. In order to obtain accurate flux densities for both bright and faint sources it is necessary to ensure that the background ``behind'' all the sources of interest is zero. After removal of large-scale background variations such as foreground cirrus, one can assume that a confusion-limited map consists of two separate populations of point sources: the ones we wish to fit, which are part of the model; and those we do not fit, which constitute the background. 
It is therefore generally true that the background is a function of the map itself (particularly the source density and beam size), and also a function of the source population that one is measuring.

\begin{figure}
\begin{center}
\includegraphics[width=0.5\textwidth]{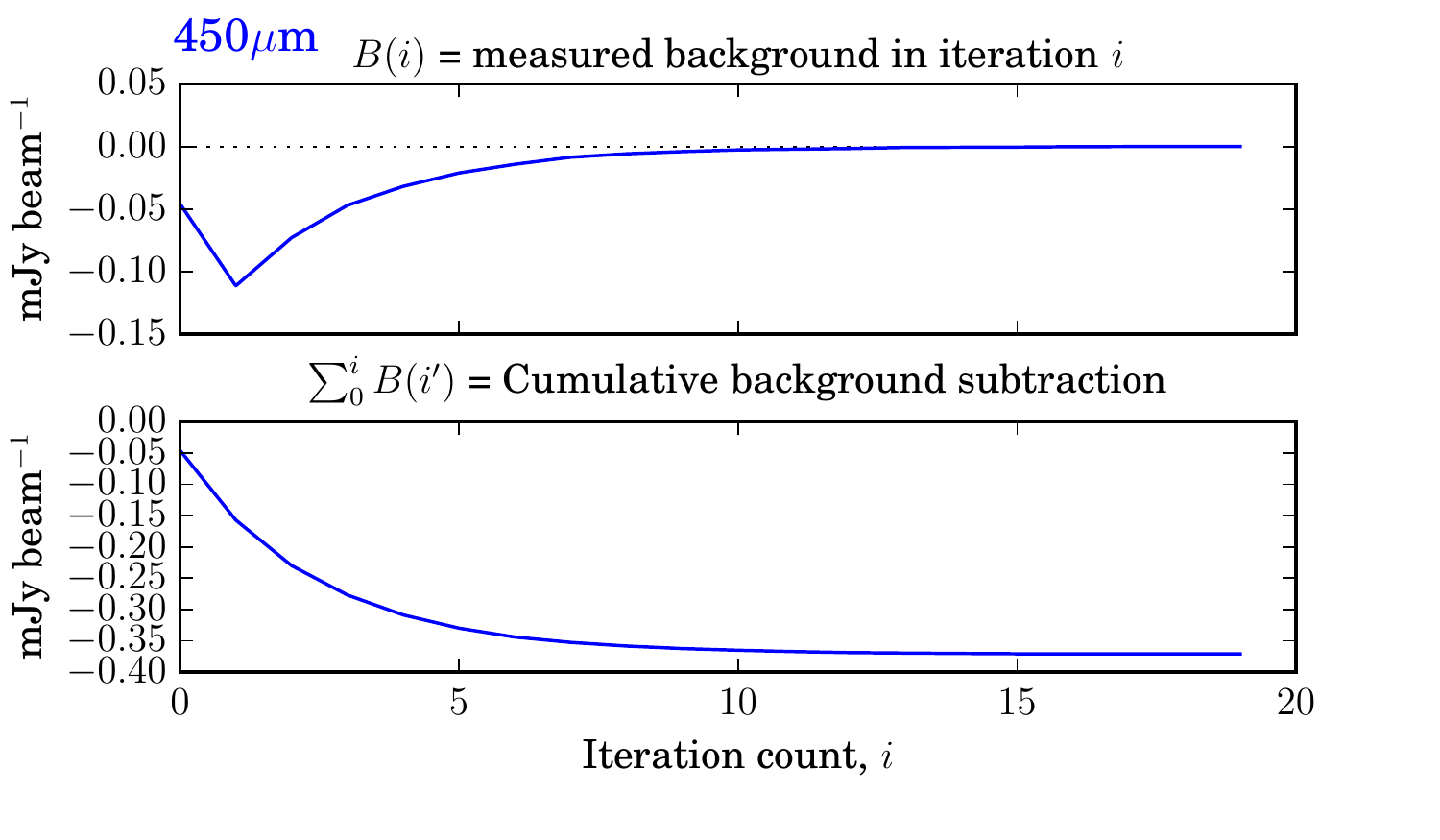}
\includegraphics[width=0.5\textwidth]{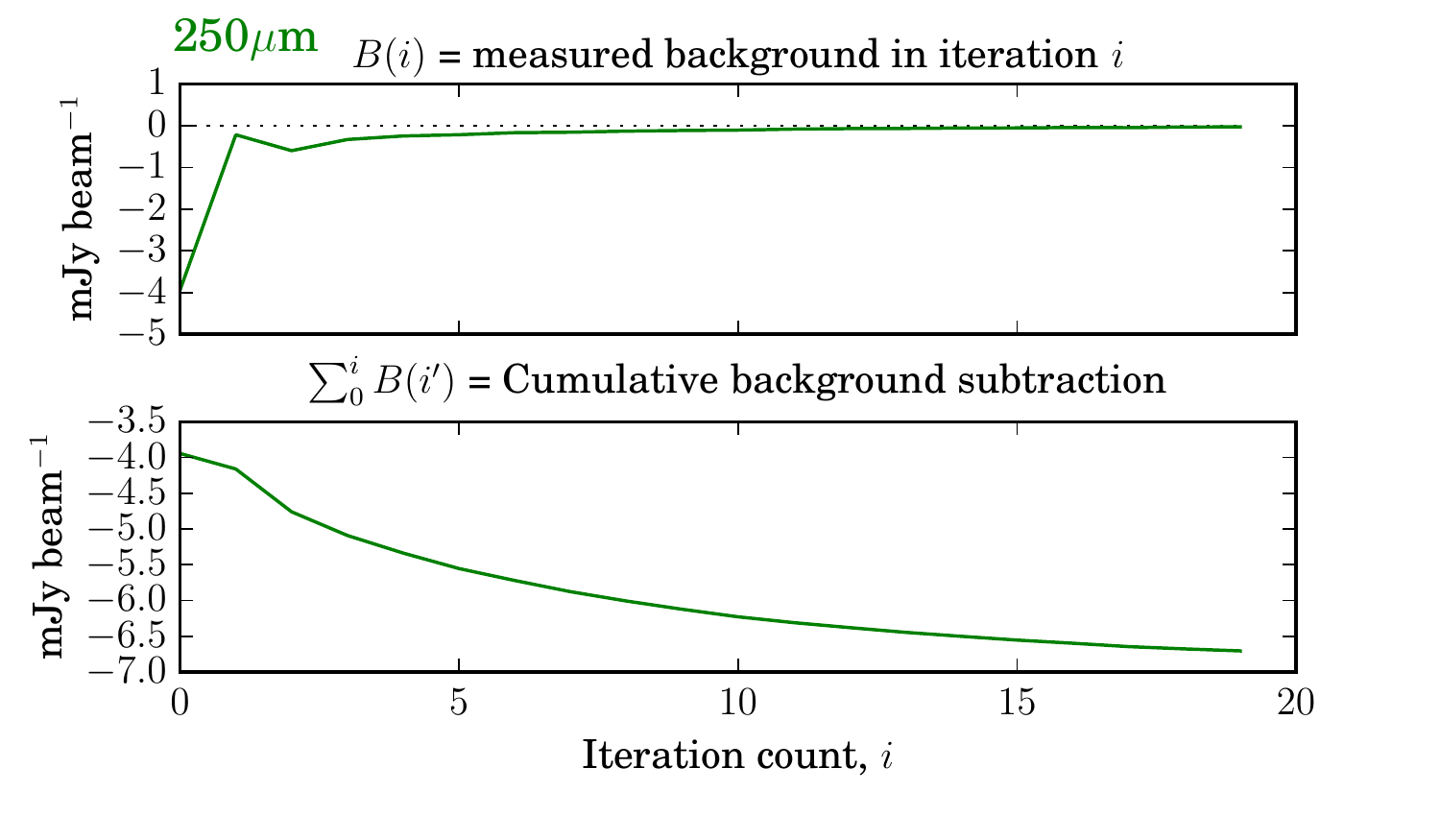}
\caption{Demonstration of convergence on the background subtraction by iterating over \tphot.
In each iteration, \tphot\ is run with the background fixed at zero; instead of fitting for the background, it is measured from the modal pixel value in the residual from the previous iteration, and is subtracted from the LRI before running \tphot. Each pair of panels show the background value $B$ measured after each iteration, and the cumulative background value subtracted from the LRI (the sum of $B$ over all previous iterations). Results are shown for the 450-\micron\ and the 250-\micron\ LRIs.
This technique gives results consistent with a single run of \tphot\ in which the background is treated as a free parameter.
}
\label{fig:bgsiter}
\end{center}
\end{figure}

One cannot fit all of the sources in the map because this would mean a greater number of degrees of freedom in the model than the number of independent data points (beams) in the data. However, one can treat the net contribution of the background sources (which are not part of the priors) as a constant background level, assuming that the variations in their surface density are uncorrelated with the sources that are in the model. 
This is effectively what is assumed in the \tphot\ background-fitting model.

As an alternative to fitting for the background as a free parameter, we can run \tphot\ with the background fixed to be zero (i.e. making the assumption that the LRI is already background-subtracted). If the background is in fact not zero, then this should result in a residual with zero mean but some skewness, and the model fluxes will be biased to balance the non-zero background. We can then iterate the following steps to converge on the best estimate for the background:
\begin{enumerate}
\item estimate the background level from the mode of the residual of the previous run (for the first run, the mode of the LRI);
\item subtract this background from the LRI used in the previous run;
\item run \tphot\ on the new background-subtracted LRI and output a new residual.
\end{enumerate}
Rather than using the mean, we estimate the mode from the centre of a Gaussian function fitted to the $2.5\sigma$-clipped histogram of pixel values in the residual map, because this is less susceptible to bias from the few bright sources that may not have been included in the model.
Convergence on the background value is generally achieved within five iterations with the 450-\micron\ LRI (and 20 iterations with 250-\micron\ LRI, which has much lower resolution), as shown in Fig.~\ref{fig:bgsiter}. If this did not converge it would indicate that too many priors have been included in the model and the fit is unstable.
The independent results of this iterative method generally agree reasonably well with the background obtained from a single run of \tphot\ with the background as a free parameter; the difference is $\lesssim0.1$~mJy~beam$^{-1}$ using the prior catalogues described above. This improves significantly when using less dense prior lists; for example using a {\it Spitzer} 24-\micron-selected prior list results in agreement within 0.01~mJy~beam$^{-1}$.

\begin{figure}
\begin{center}
\includegraphics[width=0.45\textwidth]{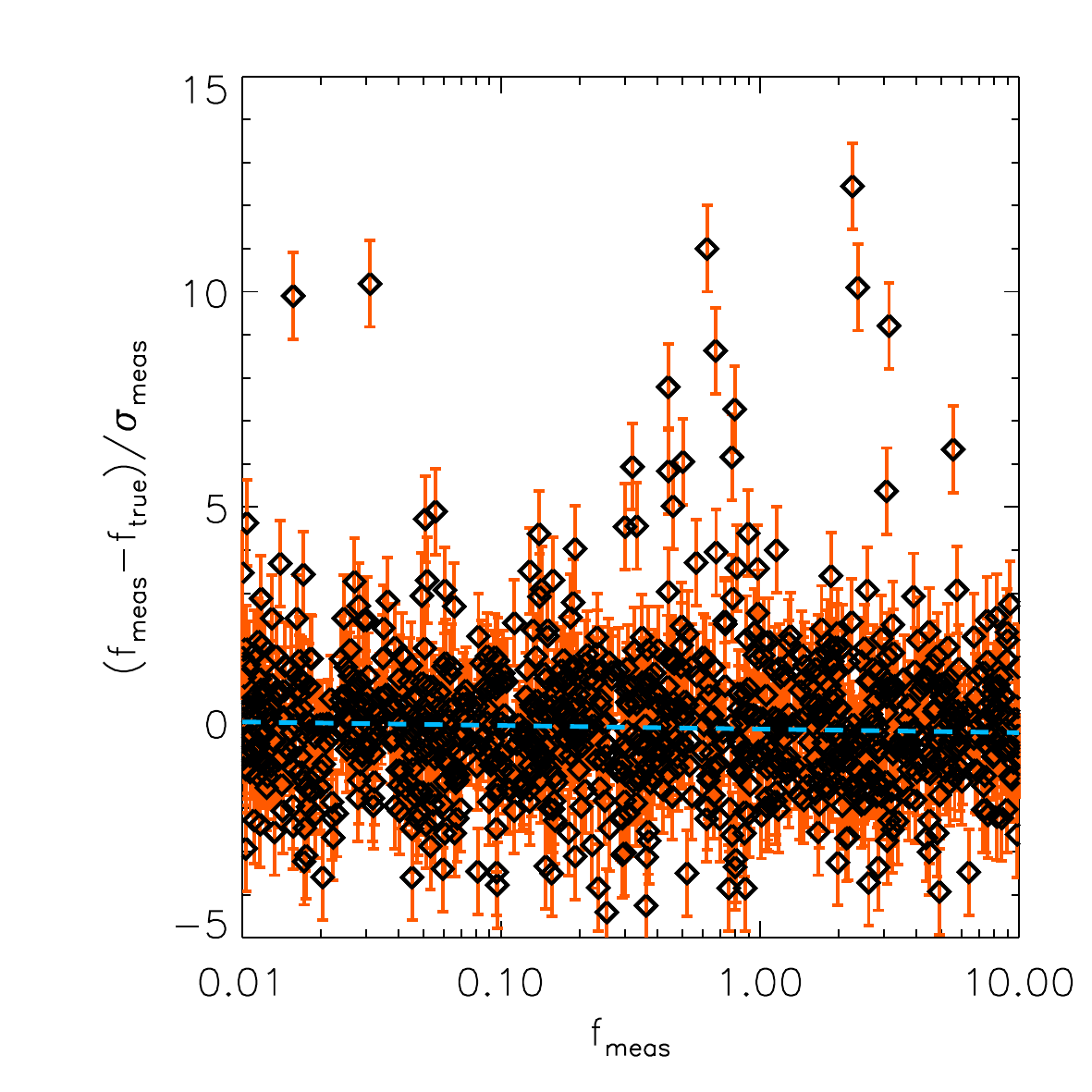}
\includegraphics[width=0.45\textwidth]{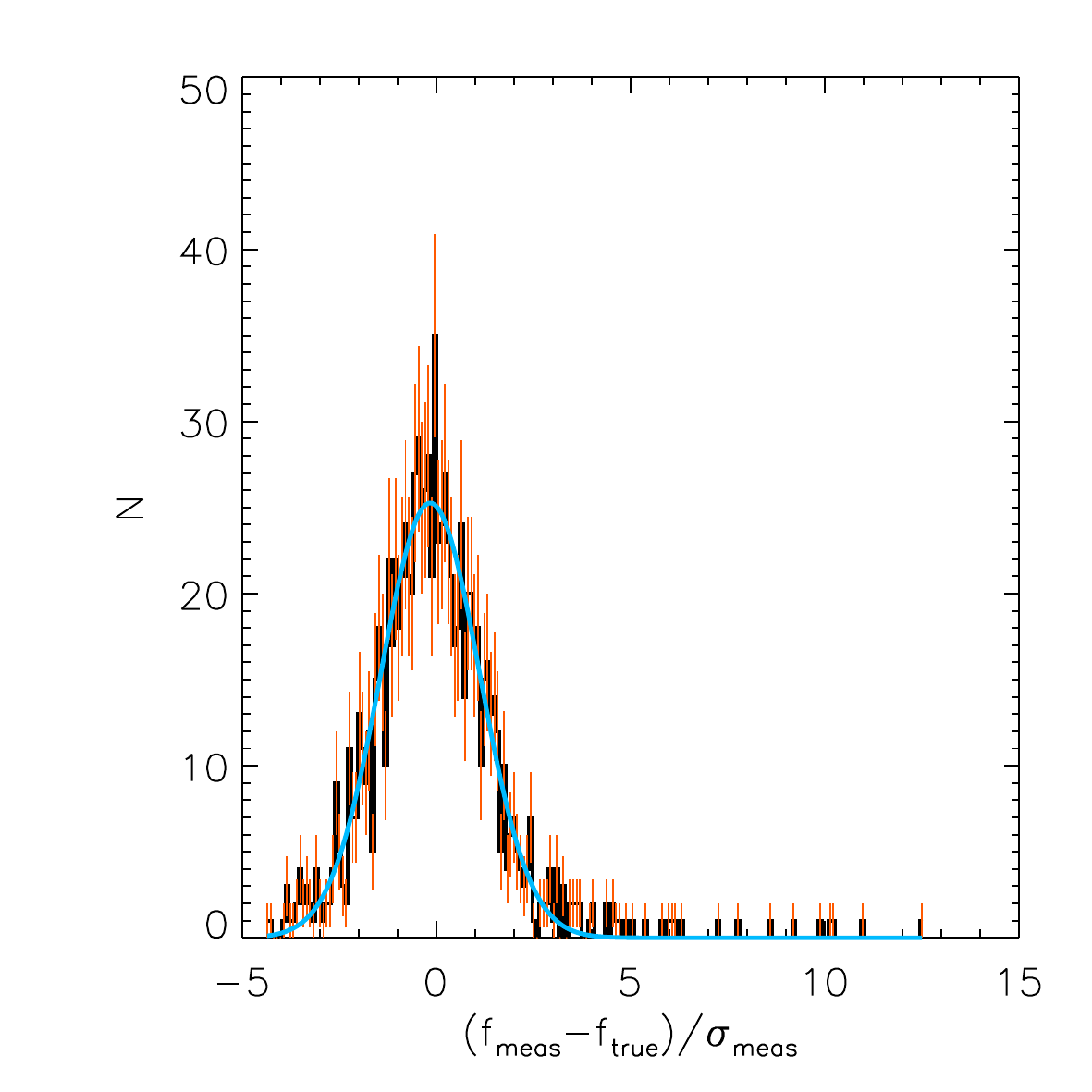}
\caption{Results of source-injection simulations to test the accuracy of \tphot\ background-subtraction. Top: the output--input flux-density difference (normalized by output error) as a function of input flux density. The linear least-squares fit shown by the dashed line is $y=(0.01\pm0.06)-(0.03\pm0.02)x$.
Bottom: the histogram of flux-density offsets, with mean of $0.15\pm0.18$ and weighted mean of $-0.02\pm0.05$. The solid line shows a least-squares Gaussian fit with mean of $-0.17\pm0.03$ and width of $1.28\pm0.03$. The simulation results indicate that output flux densities are unbiased at all flux levels.}

\label{fig:bgs_si}
\end{center}
\end{figure}

We also ran source-injection simulations in order to estimate any remaining background not captured by \tphot's internal background fitting. This method provides an estimate of the average difference between output and input fluxes of simulated sources added to the LRI. We simulated 1000 realizations, in each case adding a single point source at a random position in the LRI, and appending its coordinates to the prior catalogue of one of our fields. The fluxes of the simulated sources were randomly drawn from a distribution that was uniform in log(flux) between 0.01 and 10~mJy. We then ran \tphot\ and allowed it to fit the background as a free parameter.
The statistics of output~--~input flux (shown in Fig.~\ref{fig:bgs_si}) are approximately Gaussian with some skewness, because most injected sources fall in faint regions of the map but a few fall on top of bright sources. There is no dependence on the input flux. The median and the $2.5\sigma$-clipped mean are both $-0.27$~mJy, but the overall mean is $+0.15$~mJy. The mean is positively biased by objects with large covariance because they were injected close to existing bright sources in the map. However, the inverse-variance-weighted mean of output--input flux is very close to zero (within 0.05~mJy), indicating that the \tphot\ background subtraction is adequate if we measure the fluxes of our samples using the weighted average.

\subsection{Completeness of the sample}
\label{sec:comp}
There are several causes of potential incompleteness in our sample, resulting from the following levels of sampling:
\begin{enumerate}
\item the parent sample used to compile the 3D-HST catalogue;
\item the magnitude-limited sample within the 3D-HST catalogue;
\item the fact that not all objects in the magnitude-limited sample will have reliable stellar mass and photo-$z$ estimates.
\end{enumerate}
These have the potential to impact on our results by introducing unwanted bias into the sample, as a function of redshift and/or stellar mass. We address the potential for bias in each stage of sampling below.

\begin{figure*}
\begin{center}
\includegraphics[width=0.9\textwidth]{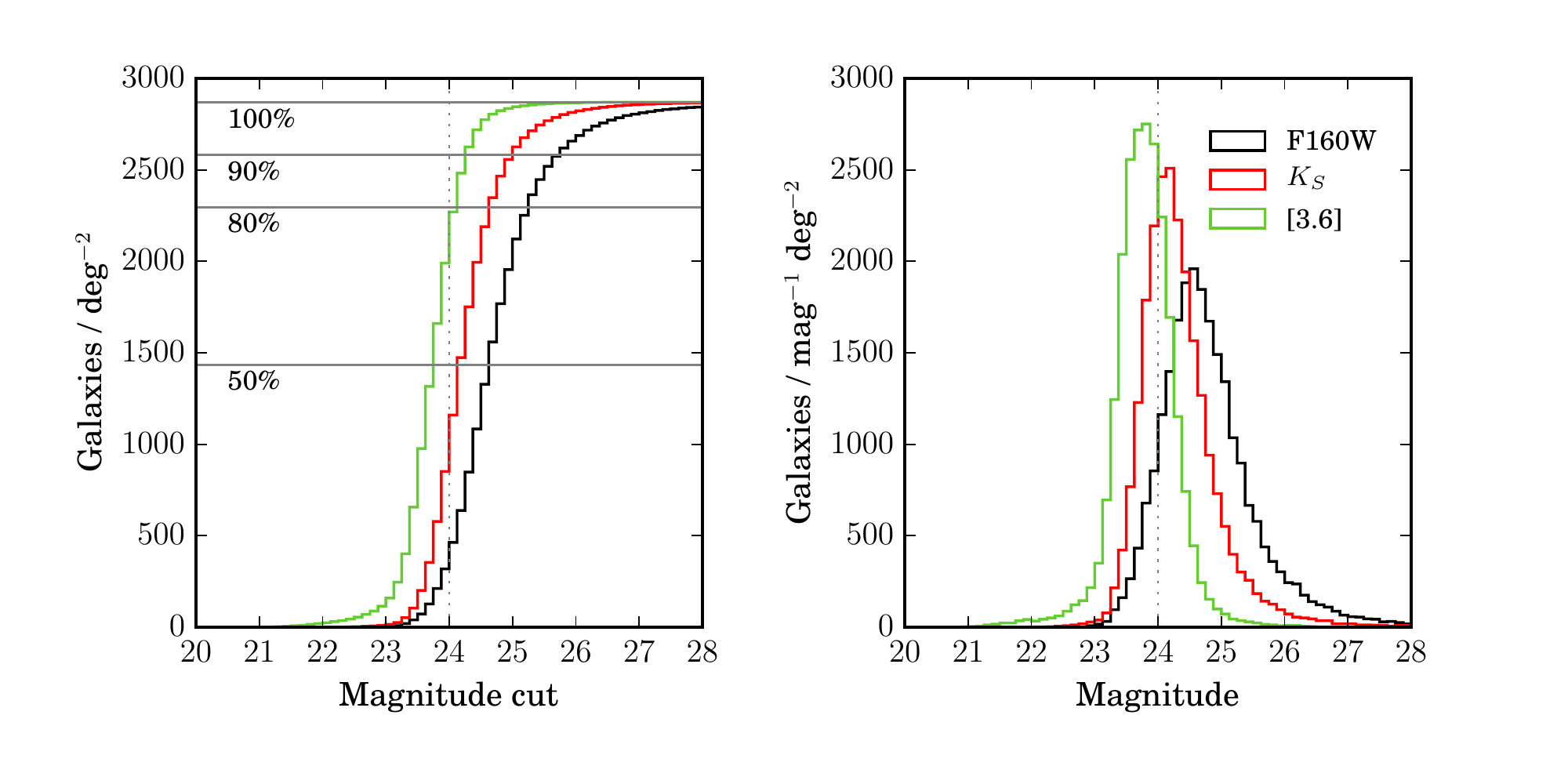} 
\caption{Completeness (left) and magnitude histograms (right) of simulated galaxies in the mass range 
$\Mstar>10^{10}\Msun$ and redshift range $4<z<6$, 
drawn from a simulated \textsc{egg} catalogue covering 10 deg$^2$ \citep{Schreiber2016}.
The three bands relevant to our selection are plotted: F160W (the effective selection band of 3D-HST, in black); $K_s$ (red); and IRAC [3.6] (green). {The dotted vertical line indicates the cut of 24 magnitudes applied in the $K$ and [3.6] bands as part of our sample selection.}
}
\label{fig:egg}
\end{center}
\end{figure*}

\begin{figure*}
\begin{center}
\includegraphics[width=0.75\textwidth]{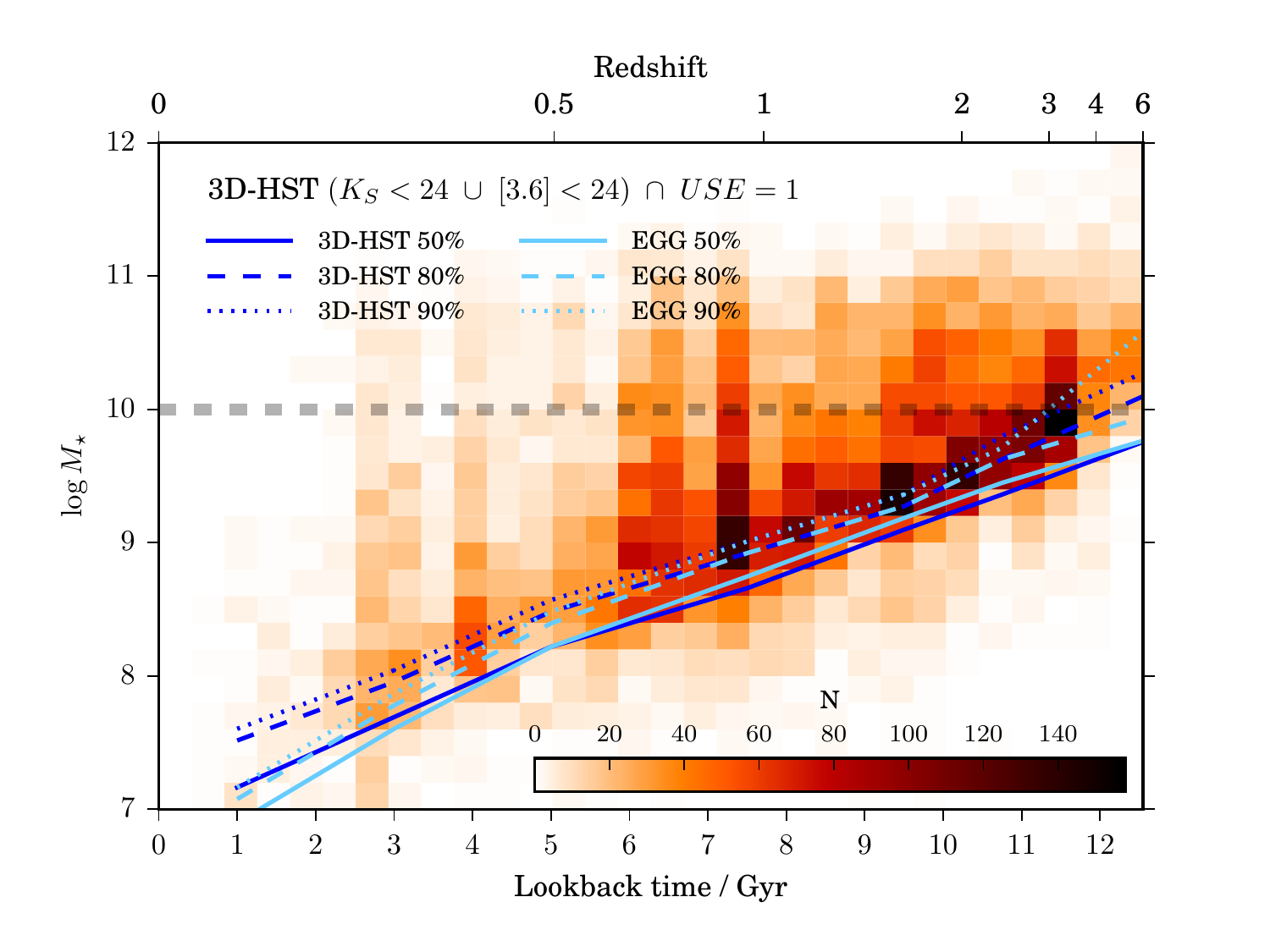}
\caption{Distribution of stellar masses as a function of redshift or lookback time, 
for the subset of the 3D-HST galaxy catalogue meeting our selection criteria (shown by the coloured density histogram). The selection criteria are given in the label. The mass limit of $10^{10}\Msun$ is shown by the {horizontal, thick-dashed} line. The three \emph{dark blue} lines mark the lowest masses to which the subset is complete at 50, 80 and 90 per cent respectively (relative to the full F160W-limited 3D-HST sample with USE=1). The three \emph{light blue} lines mark the 50, 80 and 90-per-cent completeness limits predicted from the \textsc{egg} simulated catalogue, 
where the completeness as a function of $z$ and $\Mstar$ is given by the fraction of galaxies that meet the same magnitude limits, $K_s<24$ or [3.6]~$<24$.
The agreement between the two sets of lines shows that the F160W limit in the 3D-HST catalogue does not reduce the completeness of our selection, and the overall completeness is at least 80 per cent at all redshifts.
}
\label{fig:M-t}
\end{center}
\end{figure*}

{The 3D-HST parent sample originates from {\it HST}/WFC3 imaging in AEGIS, COSMOS and UDS, and is essentially $H$-band limited with 50-per-cent completeness at F160W~$=26.5$, and 95-per-cent completeness at F160W~$=25.1$ \citep{Skelton2014}.  
From this parent sample we apply further selection constraints of $K_s<24$ or [3.6]~$<24$ (in addition to the USE flag, discussed below), which effectively select on stellar mass over the full redshift range $0<z<6$. These combined criteria are complete to a stellar mass limit of $10^{10}\Msun$ or lower at $z<4$, but will introduce further incompleteness at higher redshifts.

Fig.~\ref{fig:M-t} shows the distribution of stellar masses as a function of lookback time (or redshift), for galaxies in the parent sample which meet our selection criteria. Dark blue lines show the limiting stellar mass at which our sample is at least 50, 80 and 90 per cent complete with respect to the parent sample. 
Above a stellar mass limit of $10^{10}\Msun$, the sample completeness is greater than 90 per cent at $z<3$, and roughly 80 per cent at $3<z<6$. 

These estimates do not take into account the incompleteness of the parent catalogue, since the F160W selection is potentially biased against objects that are faint in the rest-frame UV at $z>4$ \citep{Chen2015}. We can estimate this incompleteness using realistic simulated galaxy catalogues from \citet{Schreiber2016}, as shown in Fig.~\ref{fig:egg}.
These indicate that 95 per cent of galaxies with $\Mstar>10^{10}\Msun$ at $4<z<6$ have F160W~$<26.5$, at which limit the parent catalogue is approximately 50-per-cent complete; and 73 per cent have F160W~$<25.1$, at which limit the parent catalogue is 95-per-cent complete.
Fig.~\ref{fig:egg} also demonstrates the importance of deep IRAC photometry for complete sampling at $z>4$:
only 33 per cent of simulated $\Mstar>10^{10}\Msun$ galaxies at $4<z<6$ satisfy the criterion $K_s<24$, but the [3.6]~$<24$ criterion increases this to 84 per cent.\footnote{This is slightly better than the estimated 70-per-cent completeness at [3.6]~$<24$ from S-CANDELS \citep{Ashby2015}, but is broadly consistent.}
We therefore conclude that our sample is sufficiently ($\gtrsim 70$ per cent) complete with respect to massive galaxies ($>10^{10}\Msun$) in the redshift range of interest ($0<z<6$).
We also estimate the limiting stellar mass as a function of redshift at which 50, 80 and 90 per cent of simulated galaxies meet our selection criteria; this is shown by the light blue lines on Fig.~\ref{fig:M-t}.
}

{Finally, we} consider the potential incompleteness of the subset of galaxies in the sample that have reliable photometric redshifts and stellar mass estimates from 3D-HST. \citet{Skelton2014} state that reliable measurements can be ensured by combining a magnitude cut (such as our $K_s$/IRAC criteria) with the USE flag (a flag indicating that photometry is sufficiently reliable to use for SED-fitting; see Section~\ref{sec:selection}).  
We therefore require the USE flag for our sample, which does not exclude a large fraction of sources even at the magnitude limit of 24.
The fraction of sources with $K_s<24$ or [3.6]~$<24$ that have USE~$=1$ is 83 per cent.
We cannot break this down by redshift or mass since these quantities are unreliable for sources with USE~$=0$; however, we note that the fraction does not decrease for fainter magnitudes, so we do not expect this incompleteness to be worse at $z>4$.

{In summary, as demonstrated in Fig.~\ref{fig:M-t}, our final sample contains at least 90 per cent of galaxies with stellar masses $>10^{10}\Msun$ at all redshifts up to $z=3$, and approximately 80 per cent at redshifts between $3<z<6$. 
}

\section{Analysis of T-PHOT outputs}
\label{sec:analysis}
\subsection{Flux density and error estimates}
\label{sec:errors}
{The outputs from \tphot\ provide measurements of the flux density and error associated with each prior position in each submm image. }
The \tphot\ error estimates are derived from the full covariance matrix and therefore account for all confusion between galaxies with $\Mstar>10^9\Msun$ that meet the magnitude limits of the selection. 
Furthermore, we measure an additional error from the rms of the \tphot\ residual map, which represents the residual confusion noise resulting from objects missing from the prior list.
This is added in quadrature to the \tphot\ error estimate to form the full uncertainty on each of our flux measurements, which is used for defining signal-to-noise ratios of detections.
A third source of error is the systematic flux calibration uncertainty of the instrument. Since this represents a systematic offset for all flux measurements, it does not contribute to errors between measurements made in the same waveband or instrument, but it does contribute to uncertainties between measurements from different instruments, and we consider this in Section~\ref{sec:sedfit}.
\subsection{450\micron-detected sources}
\begin{figure*}
\begin{center}
\includegraphics[width=0.45\textwidth]{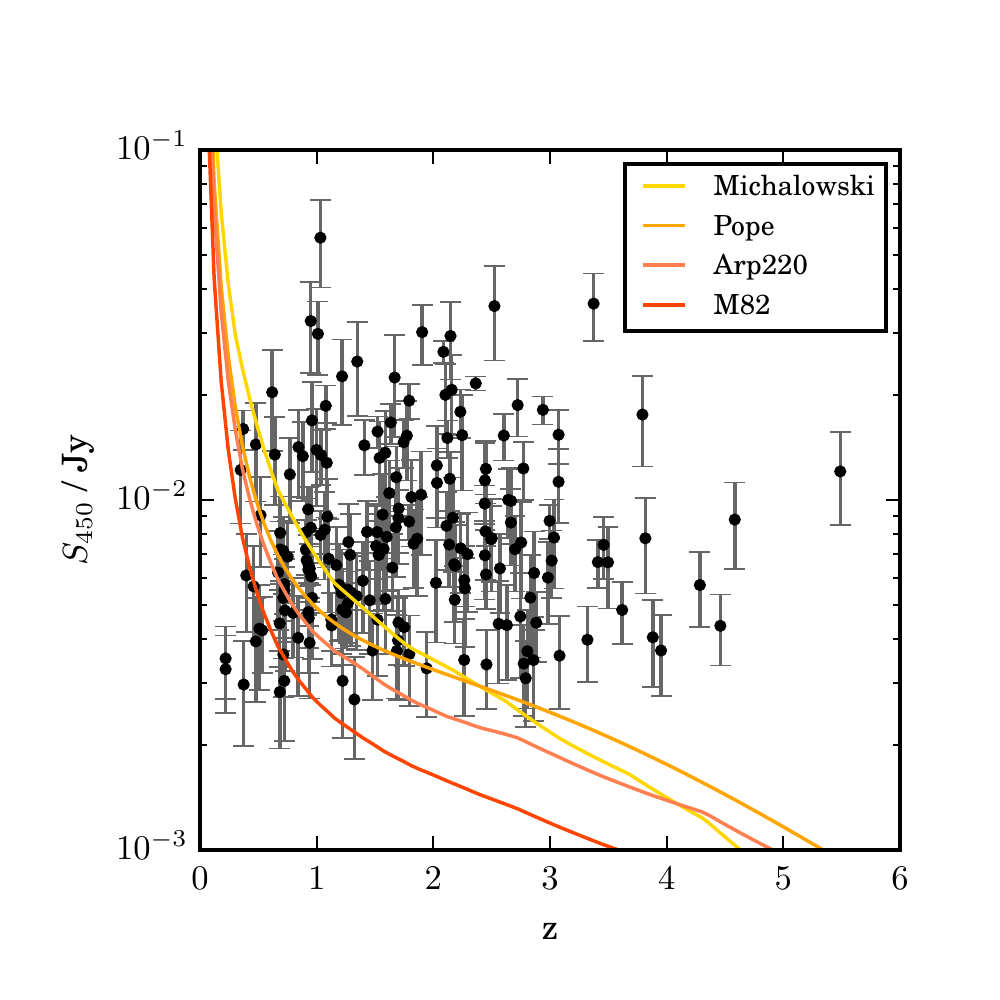}
\includegraphics[width=0.45\textwidth]{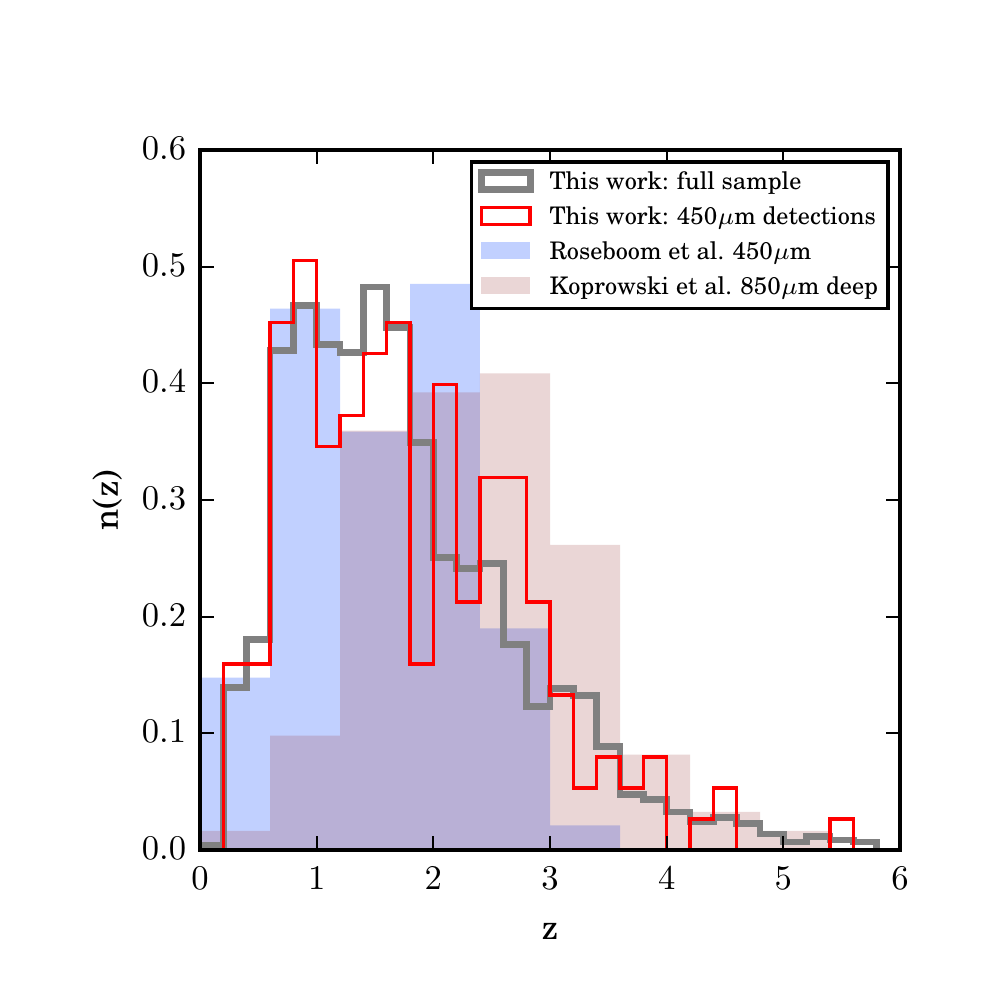}
\caption{Properties of 450-\micron\ sources detected at $S/N\geq3$ by the \tphot\ algorithm. Left: 450-\micron\ flux as a function of the redshift of the prior. Solid lines show the tracks of various SED templates scaled to a constant SFR$_\text{IR}=100\,\sfr$ ($\Lir=6.7\times10^{11}\,\Lsun$); templates are for SMGs from \citet{Michalowski2009} and \citet{Pope2008}, and for Arp220 and M82 \citep{Silva1998}. 
Right: {normalized} redshift distribution of detections (red line) and of the full prior sample with $\Mstar>10^{9}\Msun$ (grey line). We also show, for comparison, the 450-\micron-selected sources (with identifications) in the COSMOS and UDS deep fields of S2CLS \citep{Roseboom2013}; and the 850-\micron-selected sources (with identifications) in the COSMOS deep field of S2CLS \citep{Koprowski2015}.
}
\label{fig:Nz}
\end{center}
\end{figure*}

The output catalogue contains sources detected down to a limit of approximately $3$~mJy at a signal/noise ratio $S/N=3$, similar to the flux limit of blind catalogues created from the same images \citep[e.g.][]{Roseboom2013}.
By using the prior catalogue we can unambiguously associate each 450-\micron\ source with the supporting multi-wavelength photometry and SED-fitting information. We detect 165 objects from our full prior list, 130 of which have $\Mstar>10^{10}\Msun$, and 66 per cent of which have spectroscopic or grism redshifts (the rest have photometric redshifts).
The \tphot\ residuals show that there are no remaining significant 450-\micron\ sources missing from our priors. 
We can therefore make a reliable estimate of the redshift distribution of 450-\micron\ sources, as shown in Fig.~\ref{fig:Nz}. 
The median redshift of our sample is 1.68.
It is evident that sources above a fixed flux limit have a broad redshift distribution between $0.5\lesssim z\lesssim3$, as a result of the negative $k$-correction. At higher redshifts, the detection rate drops as the 450-\micron\ band probes rest-frame wavelengths blue-ward of the SED peak of star-forming galaxies, at which point the $k$-correction is less favourable. 
Nevertheless, as Fig.~\ref{fig:Nz} shows, our detections do include a small fraction of sources at $z>3$ which are largely absent from the 450-\micron-selected catalogue of \citet{Roseboom2013}. This can be explained by the difficulty of obtaining unambiguous identifications for submm sources extracted blindly from the image \citep[see also][]{Casey2013}. A similar high-redshift tail can be seen the 850-\micron-selected sample of \citet{Koprowski2015}, in which many of the sources at $z>3$ had their redshifts estimated from the 100--850\micron+1.4GHz SED, due to the absence of secure near-IR counterparts.

\subsection{Stacked results}
\label{sec:stacking}
The key advantage of our technique is the ability to estimate unbiased fluxes for sources below the confusion limit, accounting for any clustering of the sources down to scales smaller than the beam.
While only a small fraction (2 per cent) of the priors have a ``detected'' 450-\micron\ flux at $S/N\geq3$, 
it is still possible to utilise results from all priors by combining them statistically in order to measure average trends of 
450-\micron\ flux as a function of properties derived from the multi-wavelength priors.
We therefore investigate ``stacked'' 450-\micron\ properties from the \tphot\ measurements, in bins of stellar mass ($\Mstar>10^{10}\Msun$), redshift ($0<z<6$) and rest-frame UV absolute magnitude $\MUV$. We use broad bins of redshift and stellar mass to maintain a sufficiently large sample in each bin, with boundaries at $z=$~0.5, 1.5, 2.5, 4.0, 6.0; $\log(\Mstar/\Msun)=$~10.0, 10.5, 11.0, 12.0.
Note that our sample is complete to stellar masses $\Mstar>10^{10}\Msun$ at all redshifts, but by including in our \tphot\ fits all galaxies selected down to a mass limit of $10^{9}\Msun$, we ensure that the covariance matrix accounts for all cross-correlations between galaxies above this limit, in order to minimise potential correlation-induced biases.
Stacked fluxes are calculated using an inverse-variance-weighted mean in order to maximise the signal-to-noise ratio of stacked measurements.
We include all binned sources in our stacks, regardless of signal-to-noise ratio, since excluding detections would lead to a bias against bright sources.
Repeating the analysis using a median instead of a variance-weighted mean does not alter any of our conclusions.

\label{sec:binning}
In dividing our sample into bins of \MUV, we wish to probe galaxies as a function of their position on the UV luminosity function (LF). In this way we hope to understand how the UV luminosity relates to both SFR and dust obscuration. Since the UV LF evolves with redshift, we select \MUV\ bins as a function of redshift. \citet{Parsa2015} investigated the evolution of the LF via the Schechter parameters $\MUV^\ast$, $\phi^\ast$ and $\alpha$ in data from a large variety of surveys spanning $0<z<8$. They showed that the evolving LF could be empirically described by allowing these parameters to evolve smoothly with redshift. We adopt their best-fitting evolution,
\begin{equation}
\MUV^\ast = (1+z)^{0.206}(-17.793+z^{0.762}),
\end{equation}
in order to define our \MUV\ bins as a function of redshift. We divide our sample into three \MUV\ bins within each redshift bin, placing the bin boundaries at $+4$, $+2$, $0$, and $-4$ relative to $\MUV^\ast(\bar{z})$ (where $\bar{z}$ is the mean redshift in the bin). The bin boundaries have been chosen to ensure similar numbers of objects in each bin, while at the same time sampling the UV LF in a consistent way at all redshifts.

\subsection{Validating the stacking of T-PHOT measurements in simulations}
In order to test whether stacked fluxes are representative, we applied our method to a realistic mock galaxy catalogue produced by the \textsc{egg} simulation tool \citep{Schreiber2016}.\footnote{Empirical Galaxy Generator (\textsc{egg}) available from \url{http://astrodeep.eu}}
We took a simulated catalogue of a 412~arcmin$^2$ region, and created a 450-\micron\ image with realistic instrumental noise (Gaussian rms=2.5~mJy\,beam$^{-1}$), injecting point sources with a Gaussian PSF (FWHM=7.5~arcsec). 
Full details of the \textsc{egg} simulations are given in \citet{Schreiber2016}; we provide a brief description here.
The mock catalogue is constructed from stellar mass functions based on data in the GOODS-S field \citep{Schreiber2015,Grazian2015} for star-forming and quiescent galaxies separately, in redshift bins between $0.3<z<7.5$, with extrapolation up to $z=11$, and down to $\Mstar=10^8\Msun$. Stellar SEDs are assigned by measuring rest-frame colours of galaxies in CANDELS GOODS-S and defining mass-and-redshift dependent sequences for star-forming and quiescent galaxies in the $U-V$, $V-J$ (hereafter, $UVJ$) colour-colour diagram. Galaxies are modeled with disc and bulge components, which are assigned stellar SEDs according to their position in the rest-frame $UVJ$ diagram, using templates from \textsc{fast} with a \citet{Bruzual2003} stellar population.
Star-forming galaxies are assigned a SFR based on the dual-mode model of \citet{Sargent2012}; ``main-sequence'' SFRs are based on stellar mass and redshift using the fits from \citet{Schreiber2015}, with 0.3~dex scatter, and a randomly-selected 3 per cent of galaxies are placed in the ``starburst' mode by enhancing their SFR by a factor of 5.24. 
Quiescent galaxies are also assigned a SFR based on their stellar mass, in order to reproduce dust emission in these galaxies.
Each galaxy's IR luminosity is based on the SFR by assuming an obscuration fraction depending on stellar mass \citep{Schreiber2015}, and the IR SED, which depends on redshift and mass, is based on the model described in \citet{Schreiber2016}. Sky positions are assigned with a built-in angular correlation function with a power-law index of $-1$ and a normalisation which depends on redshift and stellar mass.

We selected a sample from the mock catalogue in an identical way to our true sample, fitted the simulated submm images with \tphot, and analysed results in an identical way.
Comparing output versus input fluxes indicates very good agreement with no bias in stacked results; a small fraction of individual flux measurements could be boosted by blending (as might be expected due to incomplete priors at the lowest stellar masses), but this does not bias average results in stacks. 
Measuring the difference between average output and input SFRs in bins of \MUV, \Mstar\ and redshift, we find that the distribution is consistent with the measurement errors: the distribution of $(\text{SFR}_\text{out}-\text{SFR}_\text{in})/\Delta\text{SFR}_\text{out}$ has a mean of 0.2 and a standard deviation of 0.7, and the scatter is uncorrelated with either \Mstar, \MUV, or redshift.
The input relationships between mass, \MUV\ and SFR are recovered with good fidelity, and the SFRD is also recovered without bias.
We can therefore trust that the results described in Section~\ref{sec:results} are valid and not subject to bias or systematic effects as a result of our method of measuring average SFR in bins.

\subsection{Infrared spectral energy distributions}
\label{sec:sedfit}
{Applying the techniques described above to the 450-\micron\ images provides good constraints on IR luminosities over a wide range of redshifts, thanks to the relatively high angular resolution of this imaging and the fact that it probes rest-frame wavelengths close to the SED peak at $1.5\lesssim z\lesssim 5$.
The images at 100, 160, 250 and 850\micron\ have poorer angular resolution (we assume Gaussian PSFs with FWHM = 9, 11, 18, 14 arcsec respectively) but nevertheless provide valuable additional information to constrain the shape of the SED and thus reduce the overall uncertainty of the IR luminosity measurement.
Furthermore, the \Herschel\ bands probe wavelengths close to the SED peak at $z<2$ while the 850\micron\ band is similarly valuable at $z>4$.
}
We have opted not to include the 350 and 500-\micron\ SPIRE bands due to the greatly increased confusion at these wavelengths, {meaning} that a much smaller number of priors can be reliably fitted, and for this particular study they do not offer significant additional information to constrain the SED shape.

\begin{figure*}
\begin{center}
\includegraphics[width=\textwidth,clip,trim=2cm 1cm 2cm 1cm]{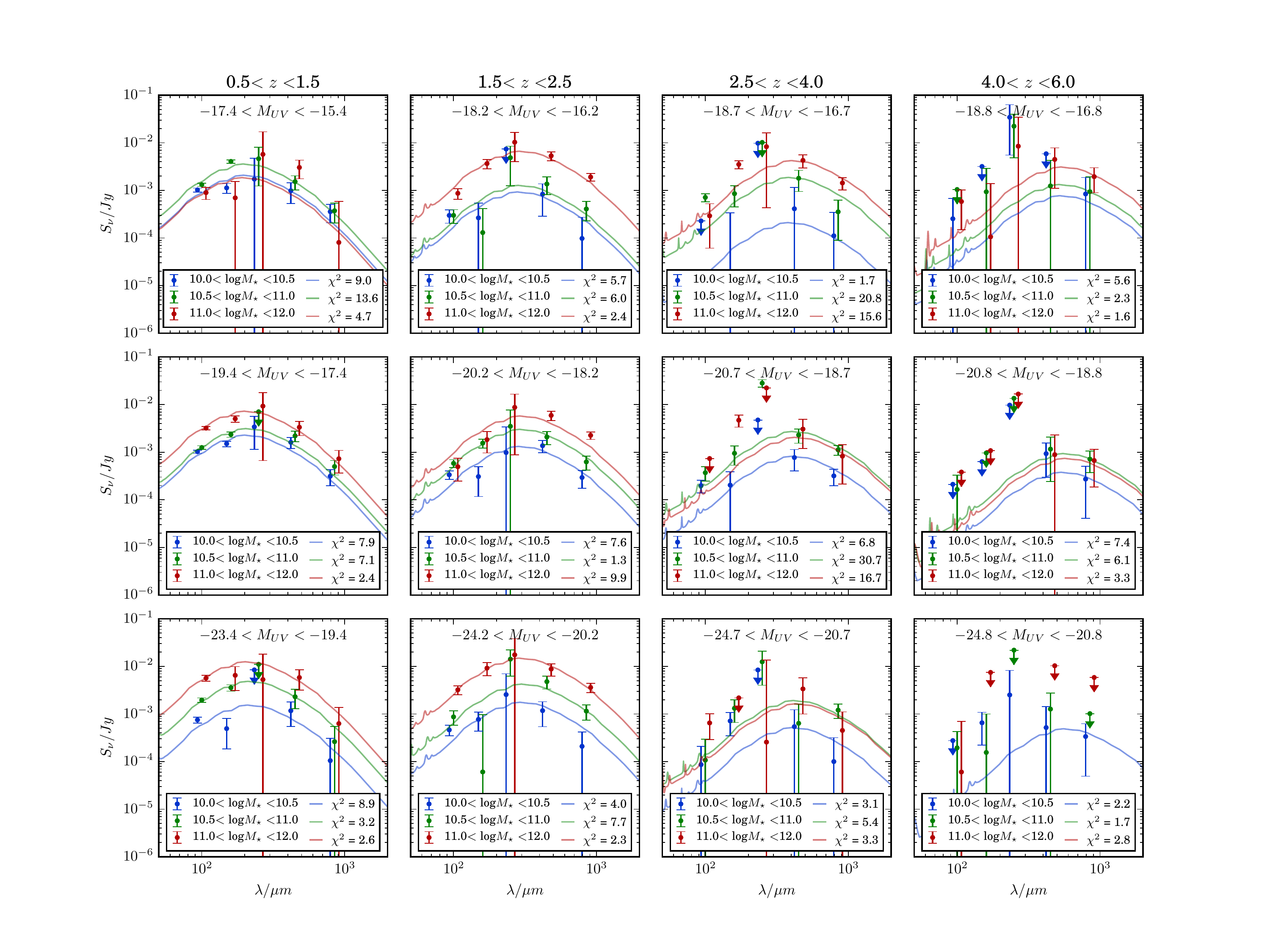}
\caption{Stacked flux measurements at 100, 160, 250, 450, 850\,\micron\ of stellar-mass- and \MUV-selected galaxies in four redshift bins (left to right) and three \MUV\ bins (top to bottom), relative to the redshift-dependent value of $\MUV^\ast$. The data points for different mass bins have been separated slightly along the $x$ axis to aid visibility.
Error bars include the \tphot\ measurement uncertainty, the residual confusion noise and the flux calibration uncertainties as detailed in the text. {Bins without a positive stacked signal are plotted as $2\sigma$ upper limits.} The best-fitting models using the \citet{Michalowski2009} SED templates are also shown, {with the $\chi^2$ for each bin shown in the legend. The $\chi^2$ takes into account the measured fluxes and errors used in the fit, and is given by \eqnref{eqn:chisq}.}
}
\label{fig:stackseds}
\end{center}
\end{figure*}

Fig.~\ref{fig:stackseds} shows the stacked flux {densities} as a function of observed-frame wavelength in each of our bins. The uncertainties on these measurements include three components.
Firstly, the error on the variance-weighted mean of the stack includes the \tphot\ measurement errors (from the diagonal of the covariance matrix, which accounts for covariances with other priors).
To each of these, we add in quadrature the residual confusion noise due to sources not accounted for in the priors, which is estimated from the standard deviation of pixels in the \tphot\ residual map.
Finally we add in quadrature the flux calibration uncertainty for each band. These are assumed to be
5.5 per cent at 100 and 160\micron,\footnote{\url{herschel.esac.esa.int/Docs/PACS/html/pacs_om.html}}
5 per cent at 250\micron,\footnote{\url{herschel.esac.esa.int/Docs/SPIRE/html/spire_om.html}}
12 per cent at 450\micron\ and 8 per cent at 850\micron\ \citep{Dempsey2013}.

In each bin, we obtained least-squares fits to several SED templates that have been previously used to describe SEDs of high-redshift star-forming galaxies. We used the average submillimetre galaxy (SMG) template from \citet{Michalowski2009}, the SMG template from \citet{Pope2008}, and the Arp220 and M82 templates from \citet{Silva1998}. 
The templates are redshifted into the observed frame at the mean redshift of each $z,\Mstar,\MUV$ bin.
{We opted to exclude the \Herschel\ data from the fitting in bins at $z>2.5$, considering that galaxies are expected to become fainter in these bands at higher redshifts. Any detection is therefore more likely to be biased by contamination from any lower-redshift objects that may be missing from the priors 
(this would explain the apparently anomalous detections at 160 and 250\micron\ in a few of the bins at $z>2.5$ in Fig.~\ref{fig:stackseds}).}
We found, however, that our results were not significantly changed if we included all data in the fitting at $z>2.5$.

Results from the fitting indicate generally very consistent SED shapes in all bins, as far as can be determined from the stacked measurements. 
The $\chi^2$ of the SED fit is given by
\begin{equation}
\chi^2=\sum\dfrac{\left(S_\text{meas}-S_\text{model}\right)^2}{\sigma_\text{meas}^2}
\label{eqn:chisq}
\end{equation}
where $S_\text{meas}$ and $\sigma_\text{meas}$ are the measured fluxes and errors used in the fit, and $S_\text{model}$ are the fluxes given by the model at the corresponding wavelengths. 
In each bin, we compared the $\chi^2$ obtained by fits to the four models, and found that the lowest $\chi^2$ was {usually obtained with the \citet{Michalowski2009} SMG template. In cases where
a different template gave the lowest $\chi^2$}, this was only a marginal improvement over the \citeauthor{Michalowski2009} template. 
We therefore decided to use the \citeauthor{Michalowski2009} {template} for fitting SEDs in all bins, {rather than allowing the SED model to vary between bins}. This template has been shown to provide a good description of the SEDs of SMGs over a wide range of stellar masses $>10^{10}\Msun$ \citep{Dunlop2016}. Furthermore, the effective cold-dust temperature of this SED (i.e. the form at $\gtrsim100\micron$) is close to a modified blackbody at 30~K with $\beta\approx1.5$, which has been shown to be an appropriate description of the SEDs of $K$-selected and colour-selected samples at high redshifts \citep[e.g.][]{Hilton2012a,Decarli2013}. 

{
Some studies have shown evidence that the average SED of star-forming galaxies evolves with redshift \citep[e.g.][]{Magdis2012,Bethermin2012b}.
Our data are insufficient to probe any such evolution accurately, due to the large error bars in many bins at high redshift. This is simply a result of high confusion noise in the \Herschel\ bands and a relatively small sample size within the area of deep 450-\micron\ coverage in S2CLS.
To test whether our results are sensitive to the assumption of a common redshift-independent SED, we first tried repeating our analysis using the template with the lowest $\chi^2$ in each bin, and found consistent results with no change to any of our conclusions.
Secondly, we tried adopting the model library of \citet{Bethermin2012b}, which predicts an evolution in the effective dust temperatures of both main-sequence and starburst galaxies due to a redshift-dependent radiation field irradiating the dust. We used this library to assign an SED template as a function of the mean redshift of each bin. The $\chi^2$ values of these model fits are very similar to those of the SMG template that we adopted, but due to the evolving temperature, these models predict slightly higher luminosities at high redshifts (by a factor of 1.2 on average at $2.5<z<4$, and a factor of 1.1 at $4<z<6$). At all redshifts, the luminosities derived from these models are within the range of the four templates described above.  Using these alternative SED models does not alter any of our conclusions, and we account for the full range of SED templates in the errors on SFRs derived from the IR luminosity (as described in the next section).
}

\subsection{Calibrating obscured and unobscured SFR}
\label{sec:sfrcalib}
We estimate the total IR luminosity (\Lir) in each bin by integrating the best-fitting model over the rest-frame wavelength range 8--1000\micron.
We estimate the uncertainty on \Lir\ by combining two components in quadrature. The first of these is the formal fitting error from the least-squares fit {to the SED}, which accounts for uncertainties in the average flux densities measured at each wavelength (see Sections~\ref{sec:errors}, \ref{sec:stacking}). Secondly, we include the full range of luminosities given by the fits to {the various SED models} described in Section~\ref{sec:sedfit}, to conservatively allow for {a wide variety of} effective dust temperatures within our bins, between cold SMG-like SEDs \citep{Michalowski2009,Pope2008}, moderate Arp220-like SEDs, and hot M82-like SEDs  {\citep[see for example][]{Magdis2012, Magnelli2013, Bethermin2015}}. 
{This SED uncertainty is generally the dominant component of the \Lir\ error at all redshifts.
The uncertainty is greatest in the $2.5<z<4$ bin. Here, the 450-\micron\ band falls very close to the redshifted SED peak, but this is where we have the poorest constraints on the SED at rest-frame $\lambda\lesssim100\micron$. Fig.~\ref{fig:stackseds} shows some apparently strong detections at 160 and 250\micron\ in a few bins at this redshift (which may be interpreted as a hotter SED). However, these are often discrepant with the measurements at 100 and 450\micron\ (which have better angular resolution). It is therefore likely that the lower-resolution \Herschel\ bands in bins at $z>2.5$ are contaminated by confusion with other sources at lower redshifts, and for this reason we have chosen to exclude the \Herschel\ data from our SED fits at $z>2.5$.
In the $z>4$ bin the two SCUBA-2 bands straddle the SED peak so the range of luminosities given by different SED models is smaller than at $2.5<z<4$.
}

We calibrate obscured SFRs by assuming the relationship 
\begin{align}
\text{SFR}_\text{IR}/\sfr&=3.88\times10^{-44}\,\Lir/\text{erg\,s}^{-1} \nonumber  \\ 
&= 1.49\times10^{-10}\,\Lir/\Lsun
\end{align}
\citep{Murphy2011}.
{We also estimate the unobscured SFR from the rest-frame far-UV (FUV) luminosity at 1600\AA, $\Luv=\nu L_{\nu,1600}$ (without extinction correction)}, which is provided in the 3D-HST catalogue based on the \textsc{eazy} templates fitted for the photometric redshifts.
The rest-frame luminosity is converted to the unobscured SFR using the relationship 
\begin{align}
\text{SFR}_\text{UV}/\sfr&=4.42\times10^{-44}\,\Luv/\text{erg\,s}^{-1} \nonumber \\
&=1.70\times10^{-10}\,\Luv/\Lsun 
\end{align}
\citep{Hao2011,Murphy2011}.\footnote{Both SFR$_\text{IR}$ and SFR$_\text{UV}$ are calibrated to a \citet{Kroupa2003} IMF.}
To estimate the \emph{total} SFR, we simply take the sum of the obscured and unobscured SFRs within a given sample, which is an established method of accurately recovering total SFR \citep[e.g.][]{Bell2003,Bell2005,Barro2011,Hao2011,Murphy2011,Davies2016}.  
Uncertainties in the measurements of \Lir\ and \Luv\ are propagated through.
{We note that the average UV and IR luminosities (and therefore SFRs) are estimated in different ways, since the UV measurement is a simple mean over all galaxies within each bin, while the IR measurement must be derived from an SED fit to stacked flux densities in multiple bands, fixed to a single redshift (the mean within each bin). We make the assumption that these two approaches are comparable assuming a reasonably narrow distribution of intrinsic luminosities within each of our mass-, \MUV- and redshift-selected bins.
}

\section{Results and discussion}
\label{sec:results}
\subsection{The dependence of SFR on stellar mass and UV luminosity}
\label{sec:sfr}

\begin{figure*}
\begin{center}
\includegraphics[width=\textwidth,clip,trim=2cm 35mm 5mm 1cm]{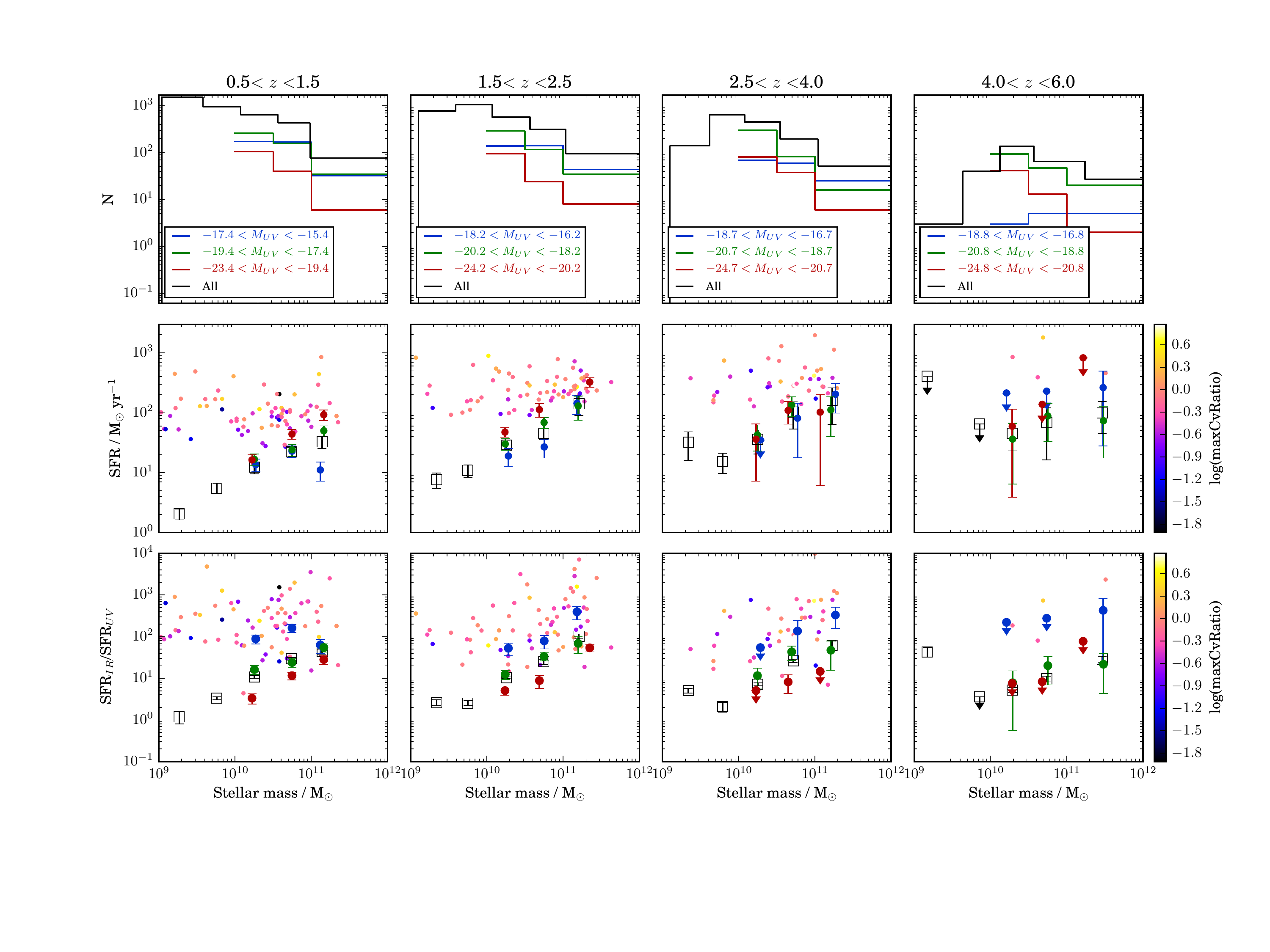}
\caption{Stacked and detected samples as a function of stellar mass for each redshift bin. From top to bottom:
number $N$ per bin; total SFR$_\text{UV}$~+~SFR$_\text{IR}$; and obscuration ratio SFR$_\text{IR}$/SFR$_\text{UV}$. 
Large black squares show the full mass-binned stacks, while large filled symbols with error bars show the stacks divided into bins of $\MUV$, defined relative to $\MUV^\ast$ at the appropriate redshift \citep{Parsa2015}.
In bins where the stacked FIR emission is not detected with $S/N>1$, the $2\sigma$ upper limit is shown as a downward arrow.
FIR detections ($S/N>3$) are shown by small coloured points in which the colour coding indicates \emph{maxCvRatio}, a \tphot\ output parameter defined as the ratio of maximum covariance to variance on the flux measurement. Individual detections with \emph{maxCvRatio}~$\gtrsim 1$ are heavily blended with another prior, which therefore dominates their uncertainty.
Note that the high SFR$_\text{IR}$ in bins with $\Mstar\sim10^{9}\Msun$ at $z>1.5$ can be attributed to incompleteness and bias of the sample at low stellar masses.  
}
\label{fig:mstar_sfr}
\end{center}
\end{figure*}

\begin{figure*}
\begin{center}
\includegraphics[width=\textwidth,clip,trim=2cm 5mm 5mm 0mm]{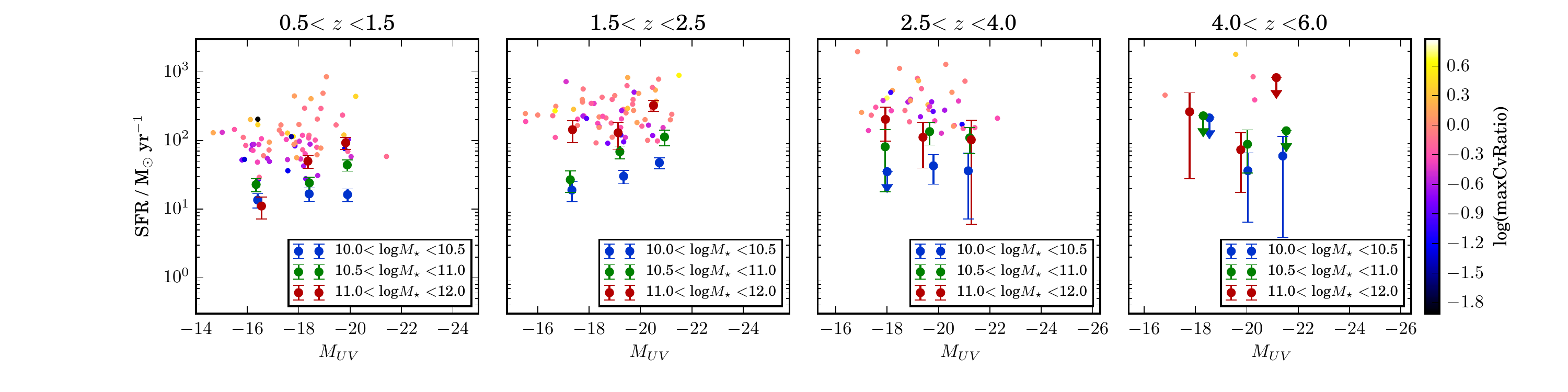}
\caption{Total SFR$_\text{UV}$~+~SFR$_\text{IR}$ plotted as a function of \MUV\ for each redshift bin. The data shown are the same as in Fig~\ref{fig:mstar_sfr}, except that colours here indicate the stellar-mass bin. 
}
\label{fig:muv_sfr}
\end{center}
\end{figure*}

{We begin by analysing our full near-IR selected sample, to explore how the average SFR varies across the range of stellar masses, UV luminosities and redshifts in the mass-complete sample.}
In Fig.~\ref{fig:mstar_sfr} we plot the total SFR measured from the stacked FIR+UV data as a function of stellar mass, divided into bins of $\MUV$ and redshift, and including all objects irrespective of measured $S/N$ (as described in Section~\ref{sec:stacking}).
The numbers of objects in each bin are shown for reference as histograms in the upper panels. 
In the lower panels, large coloured symbols show the mass/\MUV-binned stacks, while large black squares show the full mass-binned stacks with no \MUV\ binning (note that we plot these stacks down to $\Mstar=10^9\Msun$, although bins below $10^{10}\Msun$ are incomplete at $z>1.5$ as shown in Fig~\ref{fig:M-t}, hence average SFRs in those bins may be biased).
Small coloured points indicate 450-\micron\ detections, whose FIR SFR is estimated by scaling the \citet{Michalowski2009} SED template to the measured 450-\micron\ flux and integrating the template in the range 8--1000\micron\ (see Section~\ref{sec:sfrcalib}).
On this and subsequent scatter plots, the detected objects are coloured according to \emph{maxCvRatio}, an output parameter from \tphot\ defined as the ratio of maximum covariance to variance on the flux measurement. This parametrizes the level of blending uncertainty on each object; individual detections with \emph{maxCvRatio}~$\gtrsim 1$ are heavily blended with another prior, although the resulting uncertainty in their flux is accounted for in their error estimate.
It is clear that some of the high measured SFRs of detected objects could be attributed to high blending, especially for priors with stellar mass $<10^{10}\Msun$; overall, however, the \tphot\ measurements account for this blending and average (stacked) results are not affected.

Fig.~\ref{fig:mstar_sfr} shows that the total stacked SFR is correlated with stellar mass, both in the full mass-limited samples and in narrow $\MUV$ bins. 
The one exception to this relationship is that galaxies with $\Mstar>10^{11}\Msun$ and $\MUV>-17.4$ at $z<1.5$ have lower SFRs than lower-mass galaxies at the same redshift, suggesting a significant fraction of passive/quenched galaxies in this bin (we will return to this issue in Section~\ref{sec:ssfr}). 
In the lower panels of Fig~\ref{fig:mstar_sfr}, the obscuration of star formation is revealed by the fraction SFR$_\text{IR}$/SFR$_\text{UV}$, which increases strongly with increasing stellar mass and with decreasing UV luminosity.
These results indicate that UV luminosity (relative to $\MUV^\ast$) {in a given stellar-mass bin} is strongly anti-correlated with dust obscuration at all redshifts and stellar masses $>10^{10}\Msun$, and are consistent with previous studies covering similar redshift ranges \citep{Heinis2014,Coppin2015}. 
{This relationship will be investigated in more detail in Section~\ref{sec:IRXMstarMuv}.}
Fig.~\ref{fig:muv_sfr} shows the stacked SFRs plotted as a function of \MUV. 
Total SFR is relatively weakly correlated with UV luminosity at $z<2.5$, rising by a factor of 2--3 over a range of 2.5 magnitudes (a factor of 10 in UV luminosity); in constrast, the SFR-mass relation is approximately linear in Fig.~\ref{fig:mstar_sfr} (echoing the well-known ``main sequence'' of star formation; \citealp[]{Noeske2007,Elbaz2011}; see also Section~\ref{sec:ssfr}).
At $z>2.5$ in Fig.~\ref{fig:muv_sfr}, there is no discernable dependence of SFR on \MUV, and the SFR in the most UV-luminous bins is consistent with the average mass-limited SFR in Fig.~\ref{fig:mstar_sfr}.

The 450-\micron-detected sample, shown by small coloured points in Figs.~\ref{fig:mstar_sfr} and \ref{fig:muv_sfr}, samples a wide range of both stellar masses and \MUV\ at all redshifts $z<4$, with a limiting SFR~$\sim100$~\sfr\ at $1.5<z<4$. 
At all redshifts, the most luminous FIR sources (represented by the 450-\micron\ detections) sample the galaxies with the highest SFRs. These galaxies tend to be highly obscured, although they have a wide range of obscuration factors spanning roughly two orders of magnitude (lower panels of Fig.~\ref{fig:mstar_sfr}). In contrast, the most luminous UV sources (red symbols) tend to sample higher-than-average SFRs at $z<2.5$ (in comparison with the mass-selected bins; black squares), while at $z>2.5$ they appear to have similar SFRs, but are biased towards the least obscured systems at all redshifts. 

\subsection{The evolution of star-forming galaxies at high redshift}
\label{sec:ssfr}

Specific SFR (SSFR~=~SFR/\Mstar) is commonly used as an indicator of the evolutionary state of individual galaxies and galaxy samples, since it is defined as the inverse of the time required to double the current stellar mass while sustaining the current SFR. 
{This is a useful quantity because galaxies with a given star-formation history generally have similar SSFRs, even though they might have very different stellar masses. At any given redshift, most star-forming galaxies form a tight ``main sequence'' in SFR as a function of stellar mass, with a scatter of around 0.3~dex and a slope close to unity, meaning that they all have broadly similar SSFRs across the full mass range \citep{Noeske2007,Elbaz2011,Whitaker2012,Speagle2014,Schreiber2015}. 
A minority of star-forming galaxies have much higher SSFRs (when seen during a starburst phase), while passive galaxies (which are especially common at high masses and low redshifts) fall well below the sequence.
The advantage of measuring SSFR (rather than SFR) in stacking studies is that, once passive galaxies are removed, the relatively narrow distribution of SSFRs in the sample is tightly constrained by a mean or median stack, even when the range of stellar masses may be large.
}

\begin{figure*}
\begin{center}
\includegraphics[width=0.9\textwidth,clip,trim=1cm 5mm 3mm 0mm]{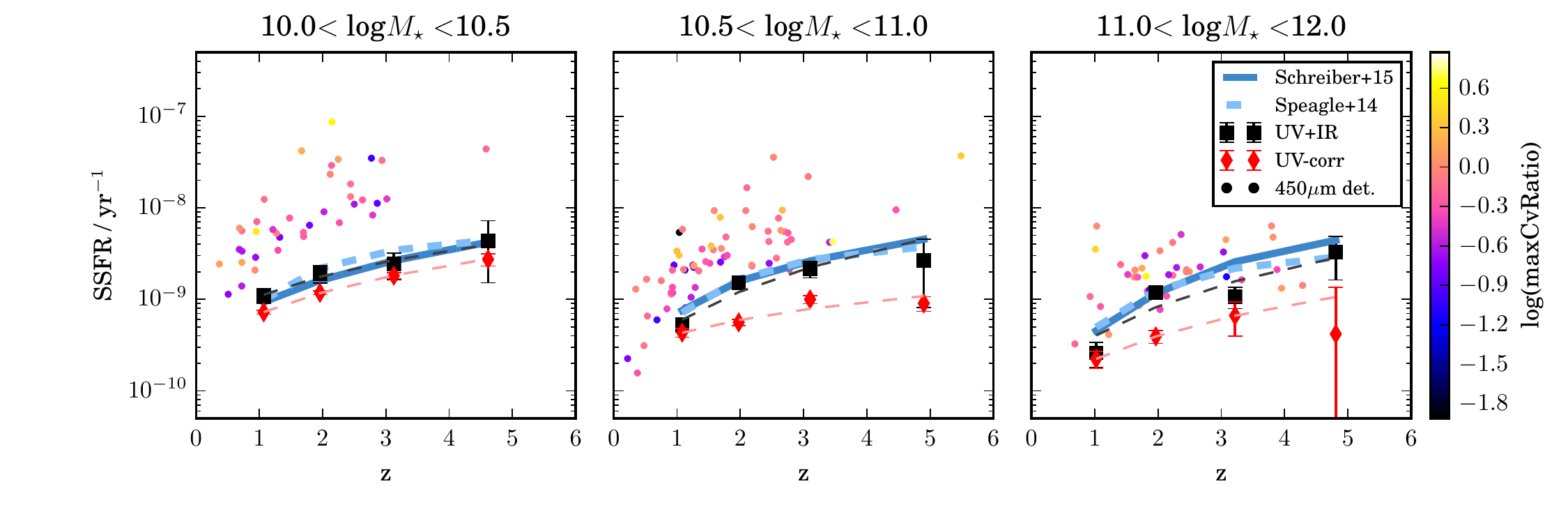}
\caption{Specific SFR as a function of redshift in bins of stellar mass, for $UVJ$-selected star-forming galaxies. Black squares are the average of the mass-limited sample, ($\Mstar>10^{10} \Msun$) using the total IR+UV SFR; red diamonds are the dust-corrected SFR estimates from the UV SED fitting alone; small points show individual 450-\micron\ detections. 
Thin dashed grey and red lines show the power-law fits to the binned UV+IR and dust-corrected UV data (respectively) as a function of $(1+z)$. 
Thick dashed and solid lines indicate fits to the main-sequence of star-forming galaxies from \citet{Speagle2014} and \citet{Schreiber2015}.}
\label{fig:z_ssfr}
\end{center}
\end{figure*}

{We therefore study the main sequence of star-forming galaxies by measuring the average SSFR as a function of redshift, shown in Fig.~\ref{fig:z_ssfr} (black squares) for the three stellar mass bins in which our sample is complete at all redshifts.
In order to restrict this analysis to star-forming galaxies, we} have excluded passive galaxies from each bin using the $UVJ$ colour criteria defined by \citet[see also \citealt{Williams2009}]{Whitaker2011}, taking the rest-frame $UVJ$ magnitudes from the SED-fitting results in the 3D-HST data release \citep{Skelton2014}.\footnote{{In Section~\ref{sec:sfr} we made no $UVJ$ cuts because we were interested in stacked SFRs across the full stellar-mass selected sample. To see the effects of these cuts on the SFR results in Figs.~\ref{fig:mstar_sfr} and \ref{fig:muv_sfr}, please see Appendix~\ref{app:uvj}.}}
The criteria for passive galaxies are
\begin{align}
\nonumber z<0.5&:\\
\nonumber U-V&>0.69+0.88(V-J),~ U-V>1.3,~ V-J<1.6\\
\nonumber 0.5<z&<1.5:\\
\nonumber U-V&>0.59+0.88(V-J),~ U-V>1.3,~ V-J<1.6\\
\nonumber 1.5<z&<2.0:\\
\nonumber U-V&>0.59+0.88(V-J),~ U-V>1.3,~ V-J<1.5\\
\nonumber z>2.0&:\\
  U-V&>0.59+0.88(V-J) ,~ U-V>1.2,~ V-J<1.4
\label{eqn:uvj}
\end{align}
Although the $UVJ$ criteria from \citet{Whitaker2011} were only defined for $z<3.5$, we extend their $2.0<z<3.5$ criteria to $z>3.5$ on the basis that massive, passive galaxies may be present even at these early times \citep{Marchesini2010,Ilbert2013, Muzzin2013,Nayyeri2014,Spitler2014,Straatman2014}.
The SSFR is calculated by taking the variance-weighted mean within each bin of all individual (SFR$_\text{UV}$+SFR$_\text{IR}$)/\Mstar\ measurements, where SFR$_\text{IR}$ is measured by scaling the 450-\micron\ flux measurement to the SMG SED template from Section~\ref{sec:sedfit}. 
{Note that this method differs from the method used to measure SFR$_\text{IR}$ from the stacked SED fit, which was combined with the mean SFR$_\text{UV}$ and mean \Mstar\ for the analysis in Section~\ref{sec:sfr}. The SED-fitting method results in smaller errors on the stacked luminosity, but the fraction $\langle \text{SFR}\rangle/\langle\Mstar\rangle$ may not be representative of the mean SSFR $\langle \text{SFR}/\Mstar\rangle$, depending on the distribution of SFR and \Mstar\ within the bin.
It is not possible to measure each individual SSFR by fitting the SED of each object (since they are not individually detected), so instead we calculate each individual SSFR from the 450-\micron\ flux and redshift, and take the average of these.}

{In this analysis} we have not binned by \MUV, since we have shown this is not a strong indicator of total SFR at a given stellar mass. 
We also show in Fig.~\ref{fig:z_ssfr} the {SFR/\Mstar\ given by} functional fits to the main sequence of star-forming galaxies, 
{SFR(\Mstar,$z$),}
from the compilation of heterogeneous results from the literature by \citet{Speagle2014}, and also the fit to stacked \Herschel\ data in GOODS, UDS and COSMOS by \citet{Schreiber2015}. The two relations broadly agree despite differing methodologies.
It is interesting to note that the 450-\micron-detected sample (small points) have SSFRs much higher than the average mass-selected galaxies at low masses, while at high masses the detected sample reaches similar SSFRs to the stacks, and is fully consistent with the main-sequence fits in the literature. Similar conclusions have previously been noted in relation to SMGs selected at similar wavelengths, from these same data-sets and others \citep{Michalowski2012a,Roseboom2013,Koprowski2014,Koprowski2015}. 

Our stacked measurements of the average UV+IR SSFR in star-forming galaxies at $\Mstar>10^{10}\Msun$ are generally consistent with the plotted relations from \citet{Speagle2014} and \citet{Schreiber2015}, as well as many other measurements not shown, covering the full redshift range 
\citep[e.g][]{Whitaker2012,Gonzalez2014,Koprowski2014,Koprowski2015,Steinhardt2014,Tasca2015}.
The average UV+IR SSFR is between $1-2$~Gyr$^{-1}$ for stellar masses $>10^{10}\Msun$ at $z\sim2-4$, and roughly $3-4$~Gyr$^{-1}$ at $z\sim5$, which is
consistent with recent submm studies at these redshifts such as \citet{Koprowski2014,Koprowski2015} from SCUBA-2, and \citet{Dunlop2016} and \citet{Schreiber2016a} from ALMA.
The evolution with redshift appears stronger at higher stellar masses, so that at $z\sim1$ there is a strong fall in SSFR with increasing mass.
This is confirmed by modelling the redshift dependence as a power law SSFR~$=a(1+z)^b$ in each mass bin, obtaining the best-fitting parameters listed in Table~\ref{tab:ssfr_fits} (fits shown by the thin grey dashed lines in Fig.~\ref{fig:z_ssfr}).
The weaker evolution of lower-mass galaxies compared with higher-mass galaxies is consistent with the results of \citet{Speagle2014} and \citet{Schreiber2015}.

\begin{figure*}
\begin{center}
\includegraphics[width=0.45\textwidth,clip,trim=0cm 0mm 5mm 5mm]{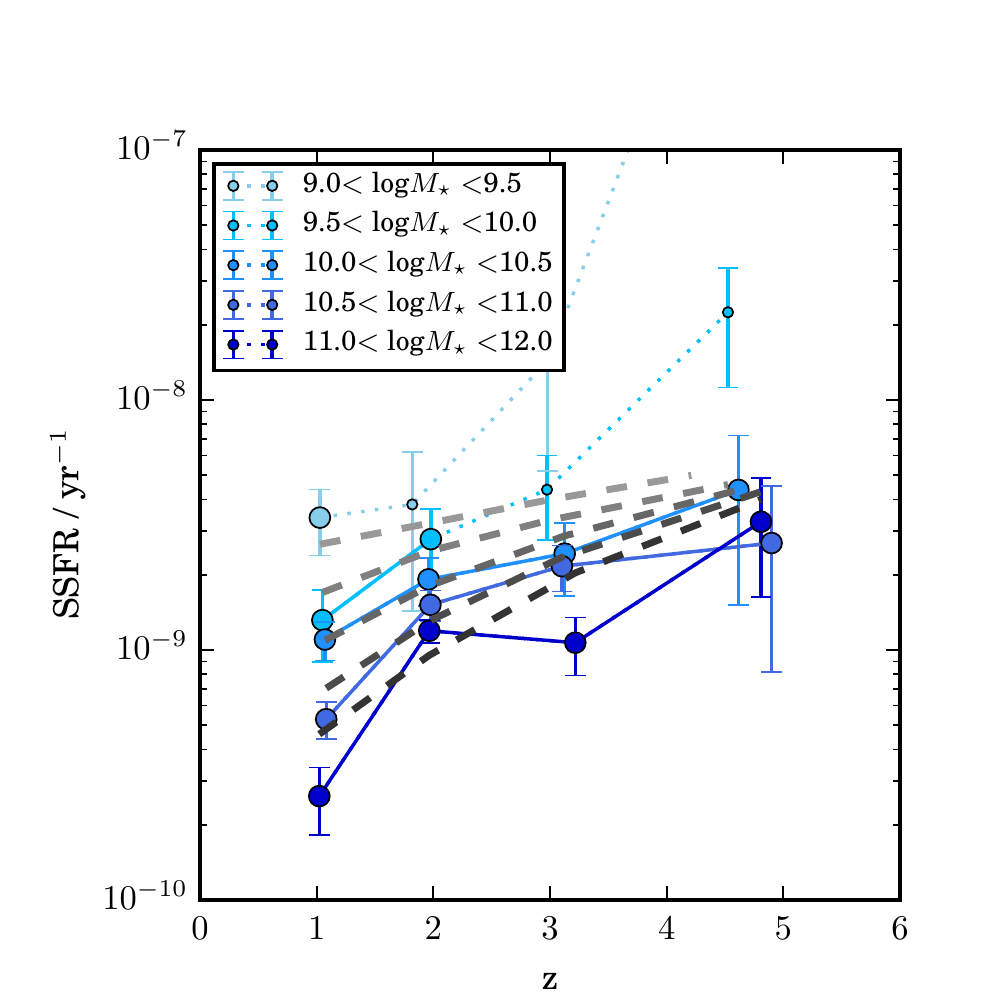}
\includegraphics[width=0.45\textwidth,clip,trim=0cm 0mm 5mm 5mm]{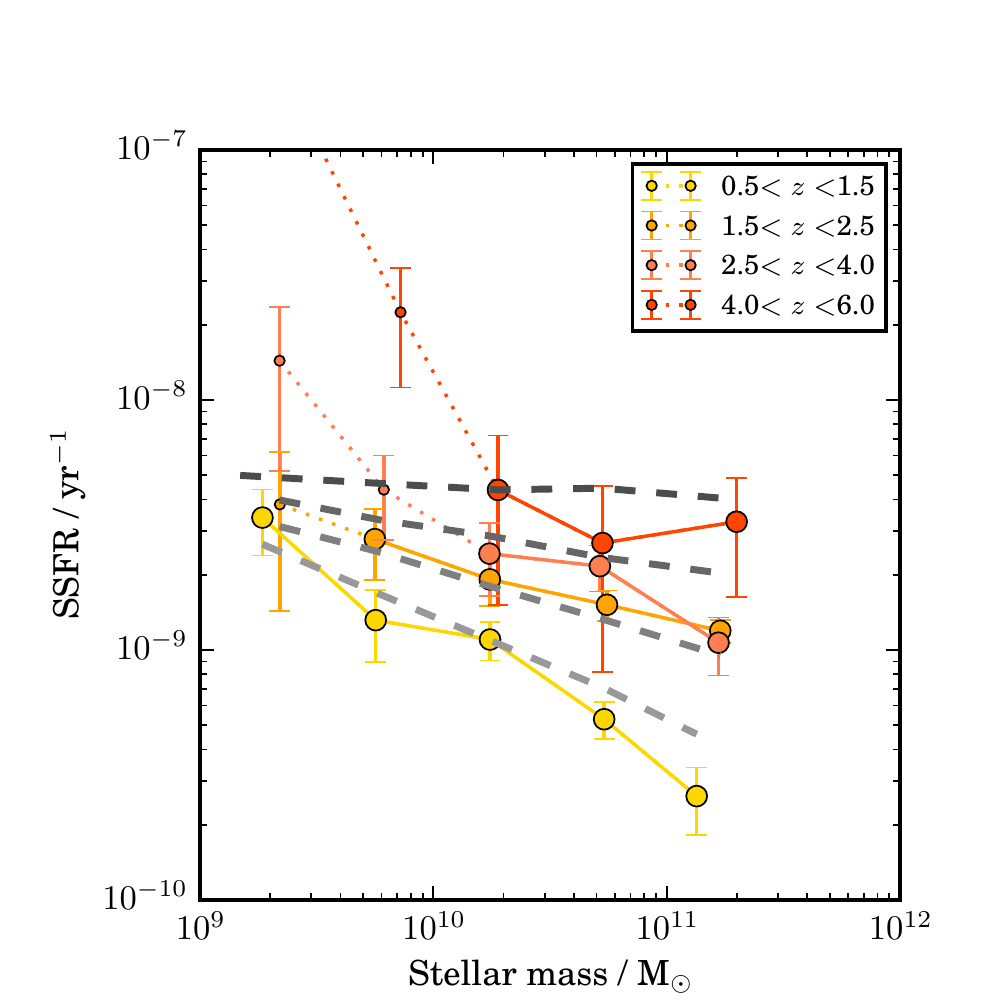}
\caption{Stacked specific SFR in bins of redshift and stellar mass, plotted as a function of redshift (left) and stellar mass (right). Large symbols joined by solid lines indicate bins in which the input sample is complete; small symbols joined with dotted lines are incomplete bins in which the SSFR is biased towards high values. Dashed lines indicate the best-fitting power-law models to SSFR($1+z$) in each stellar mass bin (left) and SSFR(\Mstar) in each redshift bin (right), fitting only to complete bins. The parameters of these models are listed in Table~\ref{tab:ssfr_fits}.}
\label{fig:m_z_ssfr}
\end{center}
\end{figure*}

\begin{table}
\caption{Best-fitting power-law parameters to the model SSFR/Gyr$^{-1}=a~(1+z)^b$ in stellar-mass bins as shown in Fig.~\ref{fig:m_z_ssfr}.}
\begin{center}
\begin{tabular}{ccc}
\multicolumn{3}{c}{IR+UV SSFR:}\\
\hline
$\log\Mstar$ & $a$ & $b$ \\
\hline
$10.0-10.5$ & $0.45\pm0.19$ & $1.3\pm0.4$ \\
$10.5-11.0$ & $0.14\pm0.05$ & $1.9\pm0.3$ \\
$11.0-12.0$ & $0.11\pm0.04$ & $1.9\pm0.4$ \\
\hline
& & \\
\multicolumn{3}{c}{UV-corrected SSFR:}\\
\hline
$\log\Mstar$ & $a$ & $b$ \\
\hline
$10.0-10.5$ & $0.27\pm0.03$ & $1.3\pm0.1$ \\
$10.5-11.0$ & $0.23\pm0.04$ & $0.9\pm0.1$ \\
$11.0-12.0$ & $0.08\pm0.04$ & $1.5\pm0.6$ \\
\hline
\end{tabular}
\end{center}
\label{tab:ssfr_fits}
\end{table}

To constrain these trends more precisely, we plot the average UV+IR SSFR as a function of redshift (left) and stellar mass (right) in Fig.~\ref{fig:m_z_ssfr}. The results in all bins are shown, although some bins are known to be incomplete (see Fig.~\ref{fig:M-t}); these are indicated by small symbols linked by dotted lines, and these bins appear to be subject to strong bias towards high SSFRs. 
The data in the complete bins appear to show a smooth dependence of SSFR on mass and redshift, which we can model as a power-law in mass with an index and normalization which both evolve with redshift \citep[e.g.][]{Whitaker2012}:
\begin{align}
\nonumber \log\left[{\rm SSFR}(\Mstar,z)/\text{Gyr}^{-1}\right]&=a(z)\left[\log(\Mstar/\Msun)-10.5\right]+b(z)\\
\nonumber a(z)&=a_0 + a_1 \log(1+z) \\
 b(z)&=b_0 + b_1 \log(1+z) 
\end{align}
The dashed lines in Fig.~\ref{fig:m_z_ssfr} show the result of a least-squares fit to the data in the complete bins (large symbols, solid lines), with the parameters 
$a_0=-0.64\pm0.19$,
$a_1=0.76\pm0.45$,
$b_0=-9.57\pm0.11$,
$b_1=1.59\pm0.24$.
The reduced $\chi^2$ of this fit is 0.8.
This model describes a negative slope in log~SSFR as a function of log~$\Mstar$ at low redshifts, which tends towards a flat relation as redshift increases towards $z\sim5$ (in other words, a sub-linear slope in log~SFR at low redshifts, which tends towards a linear relation at $z\sim5$. Similar conclusions were drawn by \citet{Schreiber2015} and \citet{Tomczak2016}, both at $0<z<4$, and \citet{Whitaker2014} at $0<z<2.5$.
In contrast, \citet{Whitaker2012} found that the slope evolved in the opposite sense over $0<z<2.5$; however, this may have been influenced by a redshift-dependent mass completeness limit in their sample. 
The slope of the main sequence at high redshifts remains a matter of active debate and much disagreement in the literature; for recent overviews see \citet{Speagle2014} and \citet{Johnston2015}.

The evolution of the normalization of the main sequence at $\log(\Mstar/\Msun)=10.5$ is described by $b(z)$, which indicates a relatively weak evolution as $(1+z)^{1.6}$, compared with some recent studies at high redshifts; e.g.
$(1+z)^{2.9}$ \citep[$0<z<2.5$;][]{Whitaker2014},
$(1+z)^{\sim2.8}$ \citep[$0<z<6$;][]{Speagle2014}, 
$(1+z)^{2.6}$ \citep[$0<z<3$;][]{Johnston2015},
$(1+z)^{\sim2.5}$ \citep[$3.5<z<6.5$;][]{Salmon2015}.
However, \citet{Schreiber2015} find the best-fitting normalization evolves as $(1+z)^{1.5}$ over $0<z<4$,
while \citet{Tasca2015} parametrize the evolution as a broken power law with $(1+z)^{2.8}$ at $z<2.3$ and $(1+z)^{1.2}$ at $2.3<z<5.5$. 
\citet{Bethermin2015} similarly suggest a break in the trend, between $(1+z)^{2.8}$ at $z<2$ and $(1+z)^{2.2}$ at $2<z<4$, while \citet{Duncan2014} fit a trend of $(1+z)^{2.1}$ at $4<z<7$.
An even weaker evolution at high redshifts is supported by \citet{Marmol2016}, who report a trend of $(1+z)^{1.0}$ based on H$\alpha$ equivalent widths at $1<z<5$.
In comparison, several theoretical models predict an evolution that closely traces the gas or dark-matter accretion rate, i.e. $(1+z)^{2.25}$, which is intermediate between the various observed trends \citep{Dutton2010a,Behroozi2013a,Guo2015}. \citet{Lehnert2015} propose a model in which SSFR rises as $(1+z)^3$ at $0<z<2$, due to an increase with redshift of gas surface densities and gas accretion rates of galaxies, but at $z>2$, self-regulating feedback maintains roughly constant (or very slowly rising) SSFRs with increasing redshift, which is consitent with some of the literature cited above. Our data suggest a moderate increase in SSFR($z$) at $z>2$, although the constraints on this slope from our data alone are relatively weak, given the large error bars and broad redshift bins.

\subsection{Obscuration and UV extinction corrections} 
\label{sec:uvcorr}

\begin{figure*}
\begin{center}
\includegraphics[width=\textwidth,clip,trim=2cm 3cm 5mm 0mm]{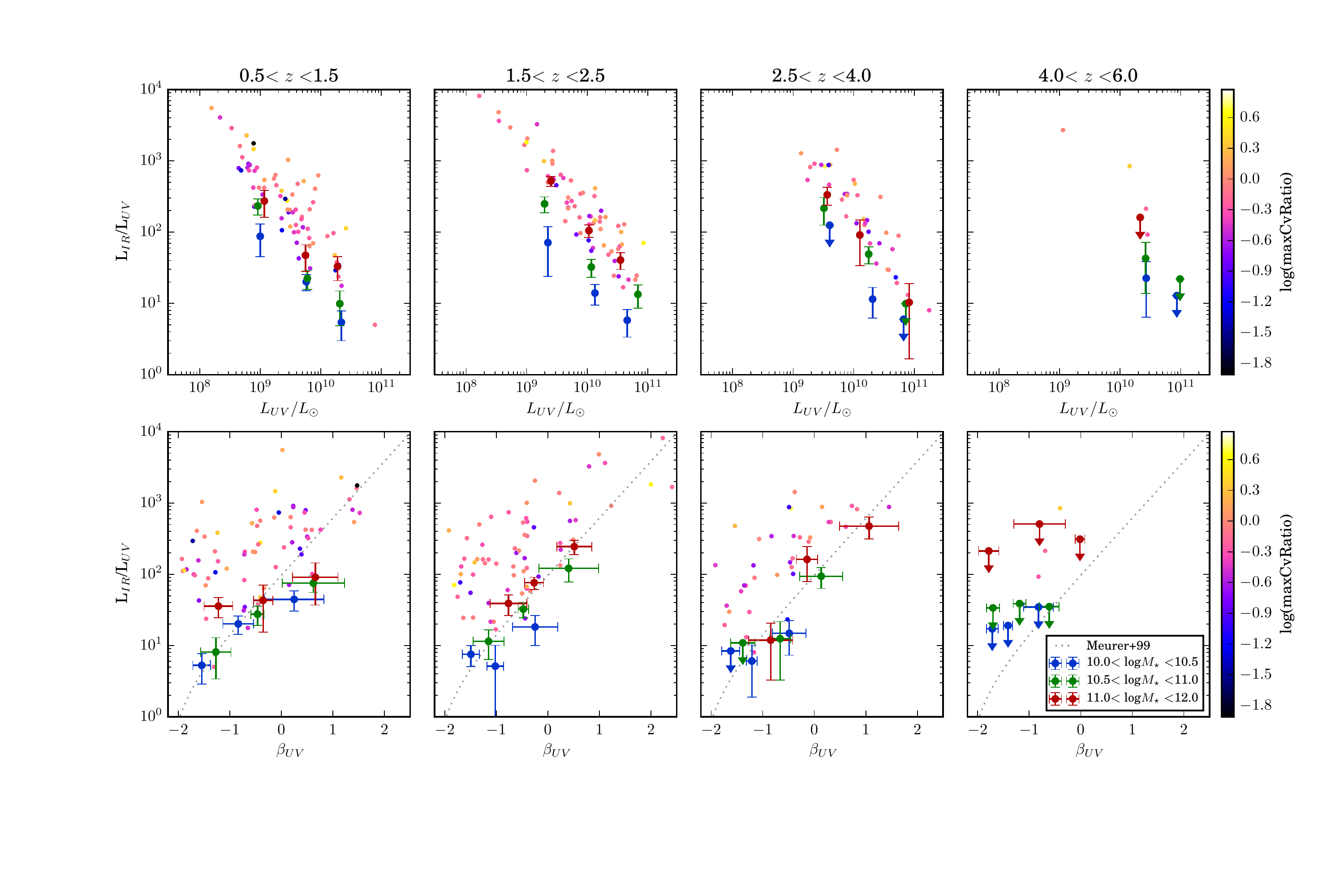}
\caption{Average IRX$=\Lir/\Luv$ as a function of UV luminosity $L_\text{UV}=\nu L_\nu(1600\AA)$ (upper panels) and UV slope $\beta$ (lower panels) from binned data and individual detections. 
{The horizontal error bars on the lower panel show the standard deviation (i.e. $\pm1\sigma$) of the distribution of $\beta$ values in each bin.}
The dotted line is the relationship for local starbursts from \citet{Meurer1999}, 
{which we have recalibrated to account for the total IR luminosity by multiplying by 1.4 (as explained in the main text).}}
\label{fig:irx_beta}
\end{center}
\end{figure*}

In Fig.~\ref{fig:z_ssfr} (red diamonds) we compare our IR+UV SSFRs to those estimated from the UV continuum luminosity following the \citet{Meurer1999} prescription {(i.e. assuming $A_\text{FUV} = 4.43+1.99\beta$)}, using the UV slope ($\beta$) fitted to the rest-frame UV data by \citet{Skelton2014}. As with the IR+UV measurements described in Section~\ref{sec:ssfr}, we plot the mean of these ``extinction-corrected'' UV SSFRs for $UVJ$-selected star-forming galaxies in each bin of stellar mass and redshift.
The results are generally lower than the ``total'' IR+UV SSFRs, and the discrepancy increases with both mass and redshift.
{This translates to a weaker redshift-evolution in SSFRs measured from the UV only, at least for $\log(\Mstar/\Msun)>10.5$, in comparison with SSFRs measured from IR+UV, leading to a lower power-law index in the fits to SSFR($1+z$) shown in Fig.~\ref{fig:z_ssfr} and Table~\ref{tab:ssfr_fits}.}

Overall these results are in much worse agreement with the literature data plotted in in Fig.~\ref{fig:z_ssfr}, indicating that these UV slope corrections may not be appropriate for stellar-mass-selected samples.
We have already seen (Fig.~\ref{fig:mstar_sfr}; Section~\ref{sec:sfr}) that the ratio SFR$_\text{IR}$/SFR$_\text{UV}$ is correlated with stellar mass and redshift, as well as being strongly anti-correlated with the UV luminosity. Since the \citet{Meurer1999} relation is calibrated for UV-bright starbursts, it is not surprising that, as our sample becomes increasingly dominated by obscured star formation at high masses and high redshifts, the UV dust corrections become less effective.

We can investigate this further by directly exploring the $\Lir/\Luv$ ratio (hereafter ``IRX'') as a function of UV luminosity [$\Luv=\nu L_\nu(1600\AA)$] and $\beta$, as shown in Fig.~\ref{fig:irx_beta}. 
Here we have again excluded passive galaxies as defined by \eqnref{eqn:uvj}. In the upper panels of Fig.~\ref{fig:irx_beta} we have binned the data by \MUV\ as in Fig.~\ref{fig:mstar_sfr} and \ref{fig:muv_sfr}, but in the lower panels we have divided the sample equally into bins of $\beta$ within each redshift and stellar-mass bin.
{Horizontal error bars show the dispersion of $\beta$ values ($\pm1$ standard deviation) within in bin.}
Stacked measurements are the average within each bin of all IRX ratios, where the individual $\Lir$ values are estimated by scaling the 450-\micron\ flux to the same SED template used above. 
{This is similar to the method used for calculating average SSFRs in Section~\ref{sec:ssfr}. 
We use this method because we cannot assume that the fraction $\langle\Lir\rangle/\langle\Luv\rangle$ is representative of the 
mean IRX, $\langle\Lir/\Luv\rangle$. The galaxies emitting most of the IR luminosity in a given bin may not be the same ones that emit most of the UV light, hence these two quantities are not equivalent. 
Because it is not possible to measure each individual IRX by fitting the SED of each object (since they are not individually detected), we instead calculate each individual IRX from the respective 450-\micron\ and UV measurements, and take the average of these.}

{In Fig.~\ref{fig:irx_beta} we also plot the IRX--$\beta$ relation of local starburst galaxies from \citet{Meurer1999}.
Since \citeauthor{Meurer1999} calibrated their relation to FIR luminosities ($L_\text{FIR}$) in the range 42--122\micron, we have applied a correction factor of 1.4 (equal to their FIR bolometric correction, i.e. the ratio of total dust luminosity to $L_\text{FIR}$) for comparison to our data.
}
In the top panels, IRX is a strong function of UV luminosity in each stellar mass bin (see also Fig.~\ref{fig:mstar_sfr}).
In the lower panels, the bins with higher $\beta$ (which in general corresponds to lower $\Luv$) generally have higher obscuration. Most of the bins are close to the \citet{Meurer1999} relation, but there is evidence for deviations in some regimes, which {may} account for the discrepancies between UV-corrected and IR+UV SSFRs in Fig.~\ref{fig:z_ssfr}.%
\footnote{{Note that Fig.~\ref{fig:z_ssfr} revealed discrepancies in the stacked SSFR (therefore SFR) calculated by the two methods in bins with $\log(\Mstar/\Msun)>10.5$, despite the fact that many bins in Fig.~\ref{fig:irx_beta} appear to be consistent with the \citet{Meurer1999} IRX--$\beta$ relation. This can be explained by a broad distribution of obscuration fractions: the average SFR in any high-mass bin contains a strong (likely dominant) contribution from highly obscured galaxies, while the \emph{mean} obscuration across all galaxies within that bin may not be extreme, and can be consistent with the mean IRX predicted from the $\beta$ distribution via the \citeauthor{Meurer1999} relation. It is therefore wise to be cautious before recommending the applicability of this relation based on stacked results.}}
At $z<1.5$ and $\Mstar>10^{11}\Msun$ the IRX--$\beta$ relation appears flatter, possibly as a result of contamination of the sample from passive galaxies (in spite of our $UVJ$ selection), although IRX is also below the \citeauthor{Meurer1999} prediction for $\beta>0$ at lower masses, which are more likely to be star-forming. At $z\sim2$ {(also perhaps $z\sim1$)} there is evidence for a stellar-mass dependence in the relationship, with $\Mstar>10^{11}\Msun$ galaxies falling above the \citeauthor{Meurer1999} relation, and $\Mstar<10^{10.5}\Msun$ galaxies falling below it.  This trend may persist at $z>2.5$, but the stacked detections are too weak in our $UVJ$-selected sample to be certain.
The dependence of this relation on stellar mass has been previously noted in several studies, using stacking of high-redshift LBGs \citep{Coppin2015,Alvarez-Marquez2016,Bouwens2016}.	
Meanwhile, FIR-detected sources in Fig.~\ref{fig:irx_beta} show a weak correlation between $\beta$ and obscuration, although most appear to have much higher obscuration than predicted by the \citeauthor{Meurer1999} relation, perhaps unsurprisingly given that the selection band is proportional to obscured SFR. The \emph{maxCvRatio} values are mostly $\ll 1$, indicating that this cannot be explained by blending-induced errors in the IR luminosities.

\subsection{Quantifying obscuration as a function of mass and UV luminosity}
\label{sec:IRXMstarMuv}
\begin{figure}
\begin{center}
\includegraphics[width=0.45\textwidth,clip,trim=5mm 0cm 5mm 0mm]{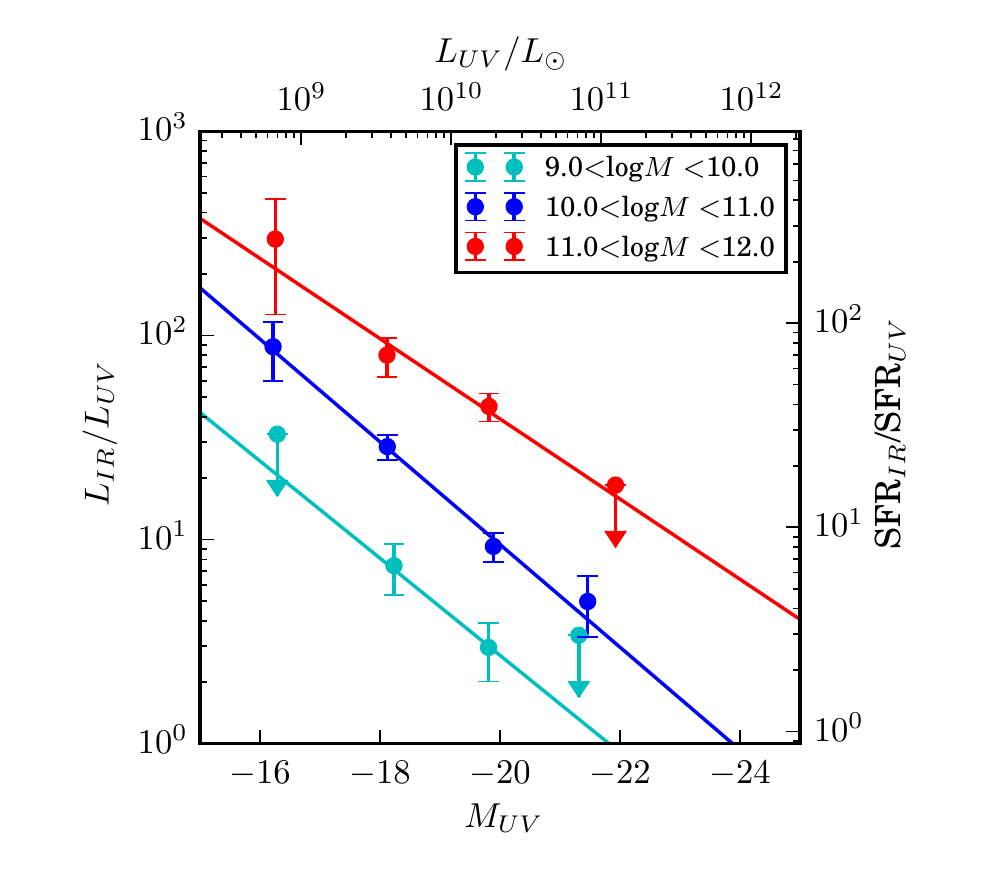}
\caption{Average IRX as a function of UV absolute magnitude in stacked data across all redshifts. Lines show the best-fitting linear models described in Table~\ref{tab:irx_fits}.}
\label{fig:irx_muv}
\end{center}
\end{figure}

The overall relationship between IRX and \Luv\ (in the upper panels of Fig.~\ref{fig:irx_beta}) appears to be independent of redshift, but strongly dependent on stellar mass. We therefore plot stacked IRX in bins of \MUV\ and stellar mass, but with no redshift binning, in Fig.~\ref{fig:irx_muv}. This shows a strong and smooth dependence of obscuration on UV luminosity in each stellar-mass bin.
Linear regression fits for $\log{(\Lir/\Luv)}$ as a function of \MUV\ in each mass bin are shown in Fig.~\ref{fig:irx_muv} and Table~\ref{tab:irx_fits}.  
The slope of this relationship is consistent in all stellar mass bins, while the normalization increases roughly as $0.5\log \Mstar$, indicating that IRX is proportional to $0.23\MUV$, or equivalently $(\Luv)^{-0.6}$; and $\Mstar^{\,0.5}$.

Alternatively, if we assume a smooth power-law dependence on both mass and luminosity, we can combine all binned data in a single model, 
\begin{equation}
\log\left(\dfrac{\Lir}{\Luv}\right) = a +b\log\left(\dfrac{\Luv}{10^9\Lsun}\right)+c\log\left(\dfrac{\Mstar}{10^9\Msun}\right)
\label{eqn:irx_mstar_muv}
\end{equation}
and find a least-squares fit with \mbox{$a=5.9\pm1.8$}, \mbox{$b=-0.56\pm0.07$}, \mbox{$c=0.70\pm0.07$} (reduced $\chi^2=0.2$).
These results are qualitatively consistent with previous \Herschel\ studies at high redshift \citep[e.g.][]{Buat2012,Hilton2012a,Heinis2014}.
{For comparison, \citet{Hilton2012a} used stacking in \Herschel\ data and found a relation $\text{SFR}_\text{IR}/\text{SFR}_\text{UV}\propto\Mstar^{0.7\pm0.2}$ in a stellar-mass-selected sample, which is equivalent to the stellar-mass dependence in \eqnref{eqn:irx_mstar_muv} above, although they did not bin by UV luminosity.
\citet{Heinis2014} stacked \Herschel\ data for rest-frame UV-selected samples at $z\sim1.5$, 3 and 4, and they also found the same stellar mass dependence, $\text{IRX}\propto\Mstar^{0.72}$, in each redshift bin.  
Binning by UV luminosity as well as stellar mass (in their Appendix A), they found that log~IRX decreases by $0.86\pm0.17$ over the range $\log(\Luv/\Lsun)=9.6$--$10.8$ (at $z=1.5$), and by $0.69\pm0.21$ over $\log(\Luv/\Lsun)=10.25$--$10.85$ (at $z=3$), equivalent to $b=-0.72\pm0.14$ ($z=1.5$) and $b=-1.15\pm0.35$ ($z=3$) in \eqnref{eqn:irx_mstar_muv}. 
This \Luv\ dependence is slightly steeper than our value of $b=-0.56\pm0.07$, although the difference is not highly significant. 
\citet{Heinis2014} concluded that there was no evidence for evolution in IRX($\Mstar$) between $z\sim1.5$ and $z\sim4$ when including all UV luminosities; however they did find evidence for evolution at a given stellar mass \emph{and} \Luv, between $z\sim1.5$ and $z\sim3$ (which they concluded was the result of an evolving \Lir--\Mstar\ relation).
Our measurement of the IRX(\Mstar,\,\Luv) relation is not divided into redshift bins (because we found no evidence for redshift dependence), and spans the broader range of $\log(\Luv/\Lsun)\approx9$--$11$. However, our sample is selected in the rest-frame optical/near-IR, and is optimised for completeness in stellar mass, therefore it is not complete in \Luv.

Finally, \citet{Whitaker2014} found a roughly linear relationship between stacked IRX and stellar mass [$9<\log(\Mstar/\Msun)<10.5$], although they did not attempt to parameterise it. However, they performed no UV binning in their stellar-mass selected sample, so it is possible that scatter resulting from the range of UV luminosities within their bins would affect the measured relationship. Furthermore, their IR luminosities were estimated from 24-\micron\ data without the benefit of longer-wavelength data close to the peak of the FIR SED.
\citet{Whitaker2014} also found that the IRX--\Mstar\ relation did not evolve over the redshift range $0.5<z<2.5$ for galaxies with $\log(\Mstar/\Msun)<10.5$, in agreement with our observations and \citet{Heinis2014}.
\footnote{{We note that \citet{Whitaker2014} found an evolution of IRX with redshift at $\log(\Mstar/\Msun)>10.5$, which was not seen at lower stellar masses, but suggested that this could be due to systematic underestimation of \Lir\ at high stellar masses, since the trend appears to conflict with low-redshift measurements of the Balmer decrement by \citet{Garn2010}.}}
}

\begin{table}
\caption{Best-fitting parameters of the linear model $\log{(\Lir/\Luv)}=p_0 + p_1(\MUV+18)$ in three stellar mass bins as shown in Fig.~\ref{fig:irx_muv}.}
\begin{center}
\begin{tabular}{cccc}
\hline
$\log\Mstar/\Msun$ & $p_0$ & $p_1$  \\
\hline
$9.0-10.0$   & $0.91\pm0.13$ & $0.24\pm0.10$  \\
$10.0-11.0$ & $1.48\pm0.05$ & $0.25\pm0.03$  \\
$11.0-12.0$ & $1.98\pm0.08$ & $0.20\pm0.05$  \\
\hline
\end{tabular}
\end{center}
\label{tab:irx_fits}
\end{table}

\subsection{Dissecting the IRX--$\beta$ relation}
\label{sec:irxbeta}

\begin{figure}
\begin{center}
\includegraphics[width=0.45\textwidth,clip,trim=2mm 0cm 2mm 2mm]{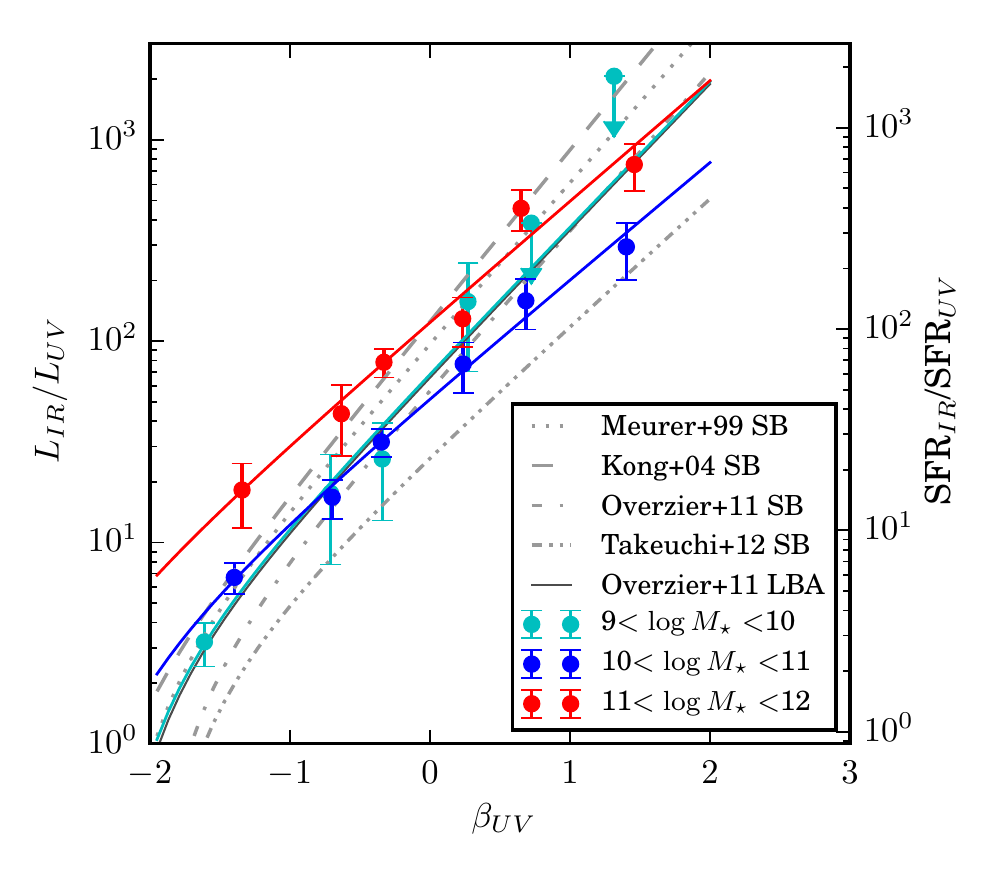}
\caption{Average IRX as a function of UV slope $\beta$ in stacked data across all redshifts, binned by stellar mass. Coloured lines show the best-fitting linear models described in Table~\ref{tab:irx_beta_fits}, {while grey} {lines show various relationships fitted to starburst samples by \citet[dotted]{Meurer1999}, \citet[dashed]{Kong2004}, \citet[dot-dash]{Overzier2011} and \citet[three-dot-dash]{Takeuchi2012}, as well as local LBG analogues from \citet[thin solid line]{Overzier2011}.}}
\label{fig:irx_beta_mstar}
\end{center}
\end{figure}

Fig.~\ref{fig:irx_beta_mstar} shows the stacked IRX--$\beta$ relation of $UVJ$-selected star-forming galaxies binned by stellar mass, but without redshift binning, assuming there is no evolution in the redshift-binned results shown in Fig~\ref{fig:irx_beta}. We now include all galaxies with stellar masses $>10^9\Msun$ (since assuming no redshift dependence, the results should be unaffected by the incompleteness at high redshifts), and can therefore better constrain the IRX--$\beta$ relationship and its dependence on stellar mass. 
{Also plotted on this figure are several functional fits to the IRX--$\beta$ relation from the literature. The first of these, the \citet{Meurer1999} relation (also shown in Fig.~\ref{fig:irx_beta}) describes the tight sequence measured for low-redshift starburst galaxies (which we have corrected to account for the total IR luminosity as described in Section~\ref{sec:uvcorr}).
Several more recent studies have re-evaluated this relationship, and we show a selection of these in Fig.~\ref{fig:irx_beta}.
\citet{Kong2004} measured a new relation for starbursts using additional \emph{IRAS} data at 12--100\micron\ and estimated total \Lir\  from an updated calibration (while \citeauthor{Meurer1999} used only the 60 and 100-\micron\ \emph{IRAS} bands and accounted for the missed fraction of IR luminosity with a constant bolometric correction of 1.4).
The \citeauthor{Kong2004} relation lies above the \citeauthor{Meurer1999} relation; but \citet{Overzier2011} and \citet{Takeuchi2012} measured relations for starbursts that fall much lower on this diagram.
Their lower values of IRX are explained by aperture effects in the UV. \citet{Overzier2011} reproduced the \citeauthor{Meurer1999} relation when considering only UV emission within a 10-arcsec radius, but when including the total UV emission of the galaxies they obtained the relation shown in Fig.~\ref{fig:irx_beta}. \citet{Takeuchi2012} came to a similar conclusion, but also showed that using new \textit{AKARI} IR data made little difference to the relation found from the \textit{IRAS} data used in the other studies.
\citet{Overzier2011} also compared the starburst sample to a sample of low-redshift LBG analogues (LBAs), measuring a relation (also shown in Fig.~\ref{fig:irx_beta_mstar}) that falls close to the original \citeauthor{Meurer1999} relation for $\beta\lesssim-1$, but closer to their updated starburst relation for $\beta\gtrsim0$.
This is particularly relevant, since the \citeauthor{Meurer1999} law is often assumed for extinction-correcting UV measurements of SFR in high-redshift LBG samples.
}

{In our stacked data in Fig.~\ref{fig:irx_beta}, we find that lower-mass ($<10^{11}\Msun$) star-forming galaxies are close to the \citet{Overzier2011} relation for LBAs, but also consistent with the \citet{Meurer1999} relation for starbursts, except in the redder bins ($\beta>0$). Galaxies with $\Mstar>10^{11}\Msun$ appear to have on average higher obscuration than predicted by either of these relations, but are perhaps closer to the relation of \citet{Kong2004}.}

Following these earlier works, we model the IRX--$\beta$ relation in each of our stellar-mass bins as 
\begin{equation}
\dfrac{\Lir}{\Luv} = \left(10^{0.4 A_\text{FUV}} -1.0 \right) \dfrac{BC_\text{FUV}}{BC_\text{IR}}
\label{eqn:irx_beta}
\end{equation}
where $A_\text{FUV}=p_0+p_1\beta$ is the attenuation at 1600\AA, and $BC_\text{FUV}$ and $BC_\text{IR}$ are the bolometric corrections in the two wavebands, taken to be 1.68 and 1.0 respectively {\citep{Overzier2011}}.
Best-fitting parameters are shown in Table~\ref{tab:irx_beta_fits}, and the fits are shown in Fig.~\ref{fig:irx_beta_mstar}. 
We note that the lowest-mass bin is consistent with the \citet{Meurer1999} relation {for starbursts} ($A_\text{FUV}=4.43+1.99\beta$) 
and is essentially identical to the \citet{Overzier2011} relation {for LBAs} ($A_\text{FUV}=4.01+1.81\beta$). 
The fit to the intermediate-mass bin is not formally consistent with either of these, although it is apparent from Fig.~\ref{fig:irx_beta_mstar} that it would be consistent if the single data point at $\beta>1$ were excluded from the fit (since we noted earlier that the reddest bins at low redshift may be contaminated by passive galaxies despite the $UVJ$ selection). The highest-mass bin ($\Mstar>10^{11}\Msun$) is inconsistent with these relations, since it is both flatter and higher in normalization: high-mass galaxies have higher obscuration than would be predicted from their UV slope, and this is particularly true for bluer UV slopes. However, as with the intermediate-mass sample, if we excluded the reddest bin from this fit we would obtain a slope similar to that of the \citeauthor{Meurer1999} relation, and the overall relation for this highest-mass bin would be similar to that of \citet{Kong2004}.

Deviations from the \citeauthor{Meurer1999} relation can be explained by several possible variables, for example star-formation history \citep{Kong2004,Casey2014}, metallicity \citep{Castellano2014}, the bolometric correction \citep{Calzetti2000} or the UV extinction curve \citep{Overzier2011,Bouwens2012,Buat2012,Salmon2015a,Safarzadeh2016}.
The \citeauthor{Meurer1999} and \citeauthor{Kong2004} relations are calibrated on local starburst galaxies, but \citeauthor{Kong2004} showed that the relation becomes less tight and has lower normalization with decreasing ratio of present/past-averaged SFR (i.e. more quiescent, less burst-dominated star formation). This may indicate that the more massive star-forming galaxies in the sample are more likely to be starbursts.
However, a lower value of $BC_\text{IR}$ {($<1$)} in more massive galaxies could also explain the deviation; this would mean that a fraction of the energy emitted in the FIR is not associated with star formation, but may be heated by older stellar populations or even AGN.

\begin{figure}
\begin{center}
\includegraphics[width=0.45\textwidth,clip,trim=2mm 0cm 2mm 2mm]{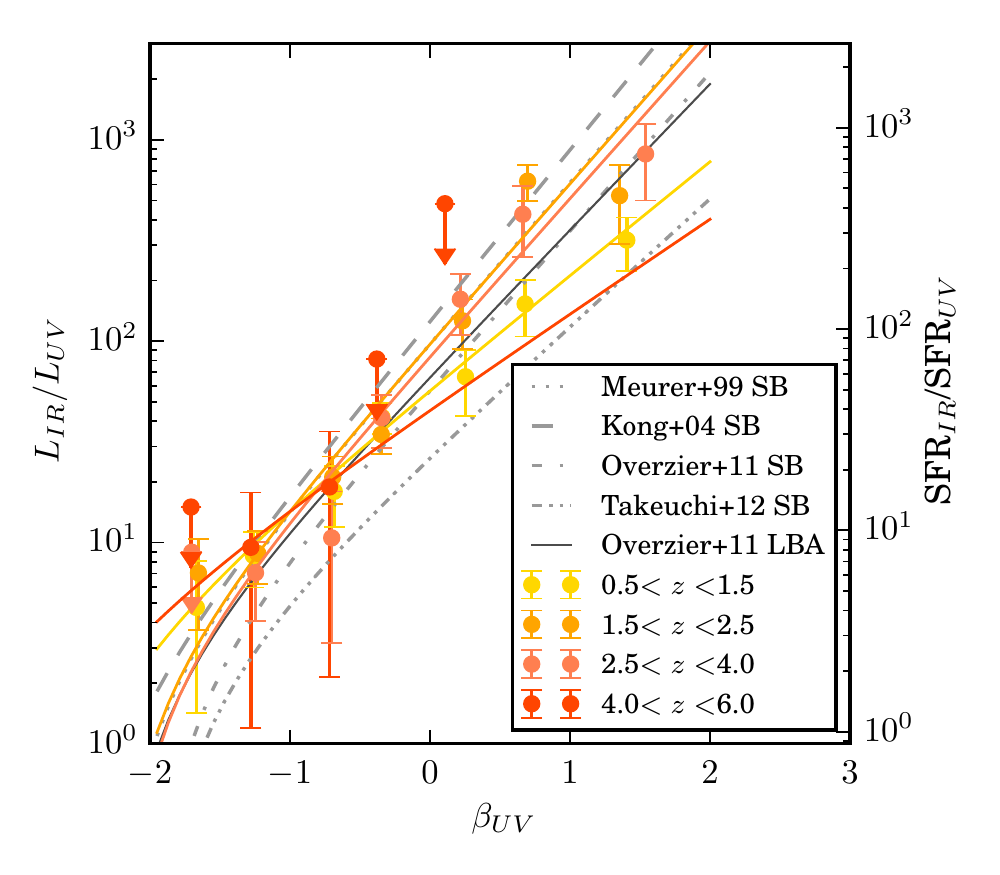}
\caption{Average IRX as a function of UV slope $\beta$ in stacked data across all stellar masses, binned by redshift. Coloured lines show the best-fitting linear models described in Table~\ref{tab:irx_beta_fits}, and grey lines are as in Fig.~\ref{fig:irx_beta_mstar}.}
\label{fig:irx_beta_z}
\end{center}
\end{figure}

\citet{Buat2012} studied the IRX--$\beta$ relationship at high redshift using \Herschel\ data, and also found that the relationship was broadly consistent with that from \citet{Overzier2011}, although with broad scatter. They did not investigate whether the deviations correlated with mass, although they did show that it correlated with \Lir, and depended on the best-fitting slope of the dust-attenuation law from their SED fitting.  
\citet{Talia2015} measured a lower and flatter relationship in a spectroscopic sample at $1<z<3$, but their sample showed a large dispersion and covered a narrow range of $\beta$, hence the uncertainties are large.
\citet{Oteo2014} used stacked \Herschel\ data to investigate the IRX--$\beta$ relationship in UV-selected galaxies at $0<z<3$, and found a strong evolution in the normalization with redshift, which is also supported by \citet{Pannella2015} in \Herschel\ stacking of a mass-selected sample at $0.5<z<4$. 
Fig.~\ref{fig:irx_beta_z} shows the IRX--$\beta$ relations that we measure as a function of redshift for all $\Mstar>10^{10}\Msun$ galaxies stacked in four redshift bins. Best-fitting parameters are listed in Table~\ref{tab:irx_beta_fits}. There does appear to be a steepening of the slope between $z<1.5$ and $z>1.5$, but there is no evidence for any evolution at higher redshifts (although the constraint at $z>4$ is very weak). We note that the flatter slope at $z<1.5$ could also be a sign of contamination of this bin by passive galaxies.  

\begin{table}
\caption{Best-fitting parameters of the model for IRX($\beta$) given by \eqnref{eqn:irx_beta}, with
$A_\text{FUV}=p_0+p_1\beta$, in the three stellar mass bins shown in Fig.~\ref{fig:irx_beta_mstar}, and the four redshift bins shown in Fig.~\ref{fig:irx_beta_z}.}
\begin{center}
\begin{tabular}{cccc}
\hline
$\log\Mstar/\Msun$ & $p_0$ & $p_1$  \\
\hline
$9.0-10.0$   & $4.0\pm0.4$ & $1.8\pm0.3$  \\
$10.0-11.0$ & $3.7\pm0.1$ & $1.5\pm0.1$  \\
$11.0-12.0$ & $4.7\pm0.1$ & $1.5\pm0.1$  \\
\hline
$z$ & $p_0$ & $p_1$  \\
\hline
$0.5-1.5$ & $3.9\pm0.1$ & $1.4\pm0.1$ \\
$1.5-2.5$ & $4.4\pm0.1$ & $2.0\pm0.1$ \\
$2.5-4.0$ & $4.3\pm0.2$ & $2.0\pm0.2$ \\
$4.0-6.0$ & $3.6\pm1.2$ & $1.2\pm1.0$ \\
\hline
\end{tabular}
\end{center}
\label{tab:irx_beta_fits}
\end{table}

Current evidence for evolution in the IRX--$\beta$ relation appears contradictory. Our results appear to suggest weak, if any, evolution in a stellar-mass-selected sample, and are consistent with results from \Herschel\ in the UV-selected samples of \citet{Buat2012} at $1<z<2$ and \citet{Reddy2012a} at $1.5<z<2.6$ [which were found to be consistent with the low-redshift relations of \citet{Overzier2011} and \citet{Meurer1999} respectively]. 
However, they do not support the strong evolution over $0<z<3$ in the UV-selected sample of \citet{Oteo2014}.
A recent study of $K$-band-selected galaxies at $1<z<3$ by \citet{Forrest2016} indicated a relationship that is higher and steeper than both \citeauthor{Meurer1999} and our own relations (\mbox{$A_\text{FUV}=5.05+2.39\beta$}, c.f. Table~\ref{tab:irx_beta_fits}), which may be attributed to differences in the analysis, in particular their use of composite SEDs fitted to individual \textit{Spitzer} and \Herschel\ FIR measurements, in contrast to our stacking technique.
Conversely, \citet{Bouwens2016} used stacking in ALMA continuum imaging and found that LBGs with $\log(\Mstar/\Msun)>9.75$ at \mbox{$z=$~2--3} lie below the \citeauthor{Meurer1999} relation, indicating an SMC-like dust extinction law, and that LBGs with lower masses lie even lower than this.
Turning to high-redshift Lyman-break-selected samples, \citet{Coppin2015} stacked \Herschel\ and SCUBA-2 data for $z\sim3$ LBGs and found that their IRX--$\beta$ relation was slightly above, but broadly in agreement with the \citeauthor{Meurer1999} relation (within the statistical error bars). They also found that LBGs with higher stellar masses were offset further above the \citeauthor{Meurer1999} relation; their high-mass bin ($\Mstar>10^{10}\Msun$) lies systematically above our data in the same mass range in Fig.~\ref{fig:irx_beta_mstar}, but is broadly consistent given the error bars.
\citet{Alvarez-Marquez2016} also used stacking of $z\sim3$ LBGs to indicate that they lie close to the low-redshift relations.
At higher redshifts, \citet{Capak2015} observed LBGs at \mbox{$z=$~5--6} with ALMA, and found that they lie well below the local relations (in a sample of 9 LBGs, selected to probe a range of UV luminosities above $L^{\ast}$), while
\citet{Schaerer2015} found that $z\sim7$ LBGs lie on or below the low-redshift relations (based on upper limits on IRX from ALMA and PdBI observations).
In contrast, \citet{Smit2016} inferred an IRX--$\beta$ relation that lies above the \citeauthor{Meurer1999} relation, and also above our own (\mbox{$A_\text{FUV}=4.98+1.99\beta$}, c.f. Table~\ref{tab:irx_beta_fits}), by comparing H$\alpha$ and UV data in a spectroscopically-selected sample at \mbox{$3.8<z<5$}. Their result is more in agreement with the results of \citet{Forrest2016}.

Overall, the results indicate a general disparity between different samples, and there is certainly a large amount of scatter in the relation at all redshifts \citep[e.g.][]{Kong2004, Overzier2011, Casey2014}, so that differences between reported correlations probably owe much to the effects of sample selection.
For example, \citet{Casey2014} showed that \Herschel-selected dusty star-forming galaxies at $0<z<3.5$ lie above the IRX--$\beta$ relation of local star-forming galaxies, with a large amount of scatter (similar to our 450-\micron-detected sample), and are therefore bluer than expected for their measured obscuration. Similar results have been discussed by \citet{Howell2010}, and the evidence points towards a dependence on the ratio of present to past-averaged star-formation rates \citep{Kong2004}, and on the effects of geometry, which can lead to decoupling between the UV- and FIR-bright regions within galaxies \citep{Howell2010,Casey2014}. Samples selected by UV luminosity, stellar mass, or Lyman-break colours may therefore show variations with mass or redshift simply as a result of differing sample biases.

\subsection{The total SFR density of massive galaxies}
\label{sec:sfrd}
\begin{figure*}
\begin{center}
\includegraphics[width=0.7\textwidth]{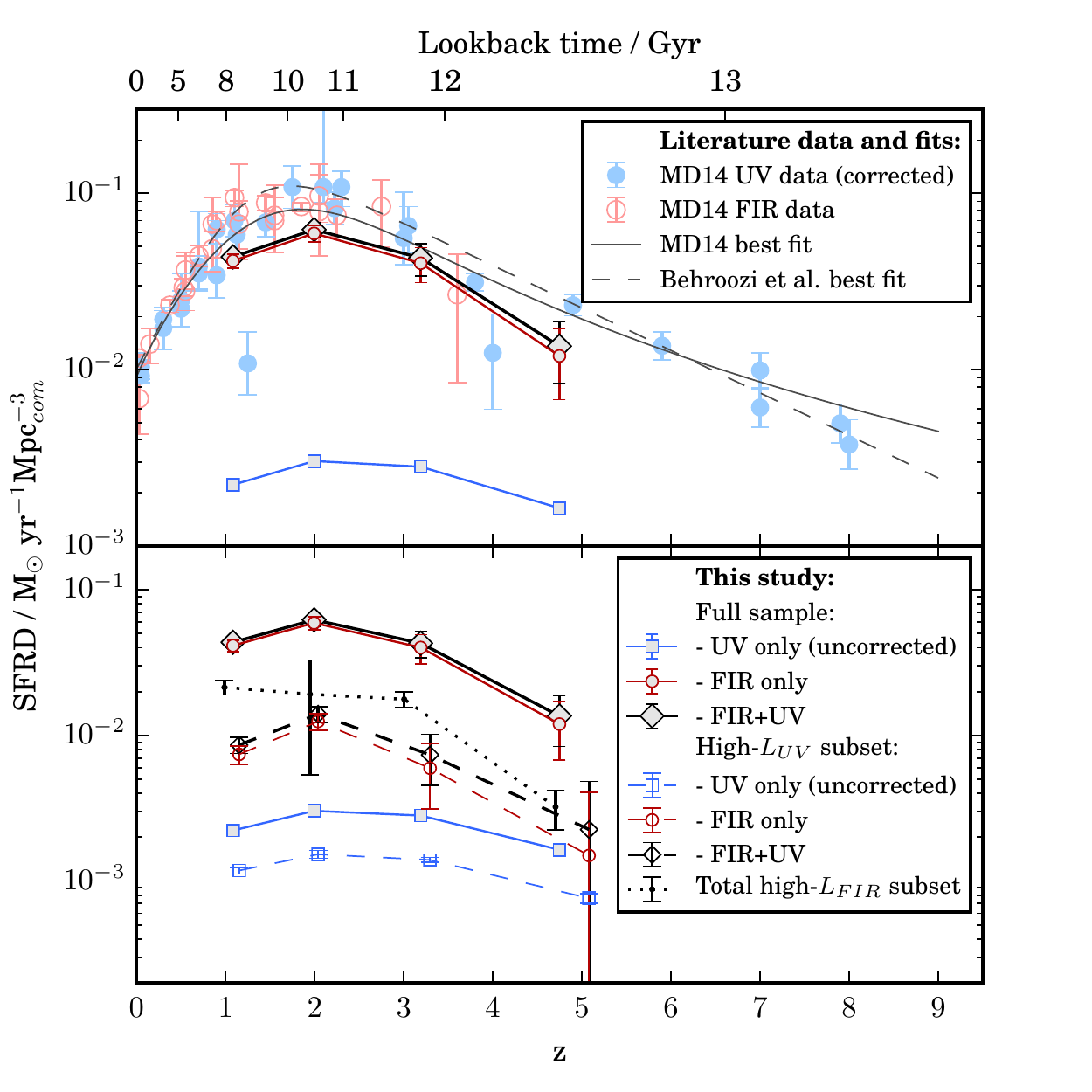}
\caption{Evolution with redshift of the comoving cosmic SFR density (SFRD) of galaxies with $\Mstar>10^{10}\Msun$, as estimated from the luminosities in the rest-frame UV (blue), FIR (red) and the total (black). The SFRD is estimated from the total SFR of all objects in a given bin of each sample. In the lower panel, these are broken down as follows: solid lines are the mass-limited sample ($\Mstar>10^{10} \Msun$); dashed are UV-luminosity-limited ($\MUV<M^{\ast}$); and dotted are detected at 450\micron\ (approximately FIR-luminosity-limited at SFR$\gtrsim 100~\sfr$).
In the upper panel, pastel coloured symbols represent results from the literature compilation by \citet{Madau2014} from rest-frame UV data (blue solid symbols) and FIR data (pink open symbols). These have been converted from a Salpeter IMF to a Kroupa IMF by multiplying by 0.61. Thin solid and dashed lines represent the best-fitting models to the data compiled by \citet{Madau2014} and by \citet{Behroozi2013b} respectively (also scaled to a Kroupa IMF).
}
\label{fig:z_sfrd}
\end{center}
\end{figure*}

We estimate the total SFR density (SFRD) of galaxies in each of the mass/luminosity/redshift bins (described in Section~\ref{sec:stacking}) by taking the sum of SFRs of galaxies in the bin and dividing by the volume of the bin, given by the difference in comoving volumes at the upper and lower redshift boundary, scaled to the sampled sky area.
{These SFRs are calculated from the stacked IR SED fit in addition to the mean UV luminosity, as described in Section~\ref{sec:sfrcalib}.}
We compute this only for galaxies with $\Mstar>10^{10}\Msun$, since the sample is complete in that range (see Section~\ref{sec:comp}), and we include all galaxies regardless of $UVJ$ colours, in order to measure the full integrated SFRD.
The results are shown in Fig.~\ref{fig:z_sfrd}, and are divided into unobscured SFRD (from raw UV luminosity; blue squares) and obscured SFRD (from {total IR} luminosity; red circles), as well as the total (large black diamonds). The data points for the full mass-selected sample are linked by solid lines, but in the lower panel of Fig.~\ref{fig:z_sfrd} we also show the contribution to the SFRD from the UV-luminous subset ($\MUV<\MUV^\ast$; dashed lines), and from the FIR-luminous subset (450-\micron\ $S/N>3$; dotted lines).

Among massive galaxies, the SFRD is dominated by obscured star formation at all redshifts by a factor of 10 or more (red versus blue solid lines). The FIR-luminous subset that is detected at 450\micron\ contributes close to a third of this SFRD (black dotted line). This roughly corresponds to sources with $S_{450}>3$~mJy and SFR~$\gtrsim100\sfr$, although the limiting SFR of this sample does evolve with redshift. 
The UV-luminous subset ($\MUV<\MUV^\ast$; black dashed line) contributes a slightly smaller fraction of around one fifth at all redshifts. 
This is partly because these objects are rare, but also because they do not generally have high SFRs compared with typical galaxies of a similar mass (see Fig.~\ref{fig:mstar_sfr}). In fact the FIR-bright galaxies (black dotted line) are much rarer, but contribute a significantly higher fraction of the total SFRD, at least at $z<4$ (see also \citealp{LeFloc'h2005,Murphy2011a,Casey2012a,Casey2013}).
Notably, even in the UV-luminous sample, the SFRD is dominated by the obscured portion (red versus blue dashed lines), although this becomes less dominant with increasing redshift.
For the FIR-detected sample, we find that the SFRD is almost completely obscured; less than one per cent of their SFRD is detected in the UV (not shown in the figure).

In the upper panel of Fig.~\ref{fig:z_sfrd} we show SFRD measurements from the literature compiled by \citet{Madau2014}; these are indicated by the light blue and pink symbols (representing UV and FIR measurements respectively). These have been recalibrated from a \citet{Salpeter1955} to a \citet{Kroupa2003} IMF for comparison with our results, by multiplying by a factor 0.61.
Our data provide a new, direct measurement of the obscured SFRD of stellar-mass-selected galaxies at $z\sim5$,
highlighting the importance of our combination of high-resolution SCUBA-2 maps and high-quality multi-wavelength prior catalogues.
Previous studies with \Herschel\ have generally reached only up to $z\lesssim4$ due to the much higher confusion limit,
either using stacking \citep[e.g.][]{Viero2013,Schreiber2015} or by extrapolating and integrating measured luminosity functions of detected sources \citep[e.g.][]{Burgarella2013a,Gruppioni2013}. The selection and identification of bright, red submm sources can provide lower limits on the obscured SFRD at $z>4$ \citep{Dowell2014,Shu2015}, but it is impossible to reliably estimate the total SFRD from these results which only probe the very bright end of the IR LF.
However, it has already been shown that SCUBA-2 can offer a view of the $z>4$ obscured SFRD in SMGs \citep{Casey2013} and stacked LBGs \citep{Coppin2015}, and we have extended this view to include a wider sample of stellar-mass-selected galaxies.
We will directly compare our results to the literature in the next section.

\subsection{Recovering the full SFR density}
\label{sec:lmc}
\begin{figure*}
\begin{center}
\includegraphics[width=0.7\textwidth]{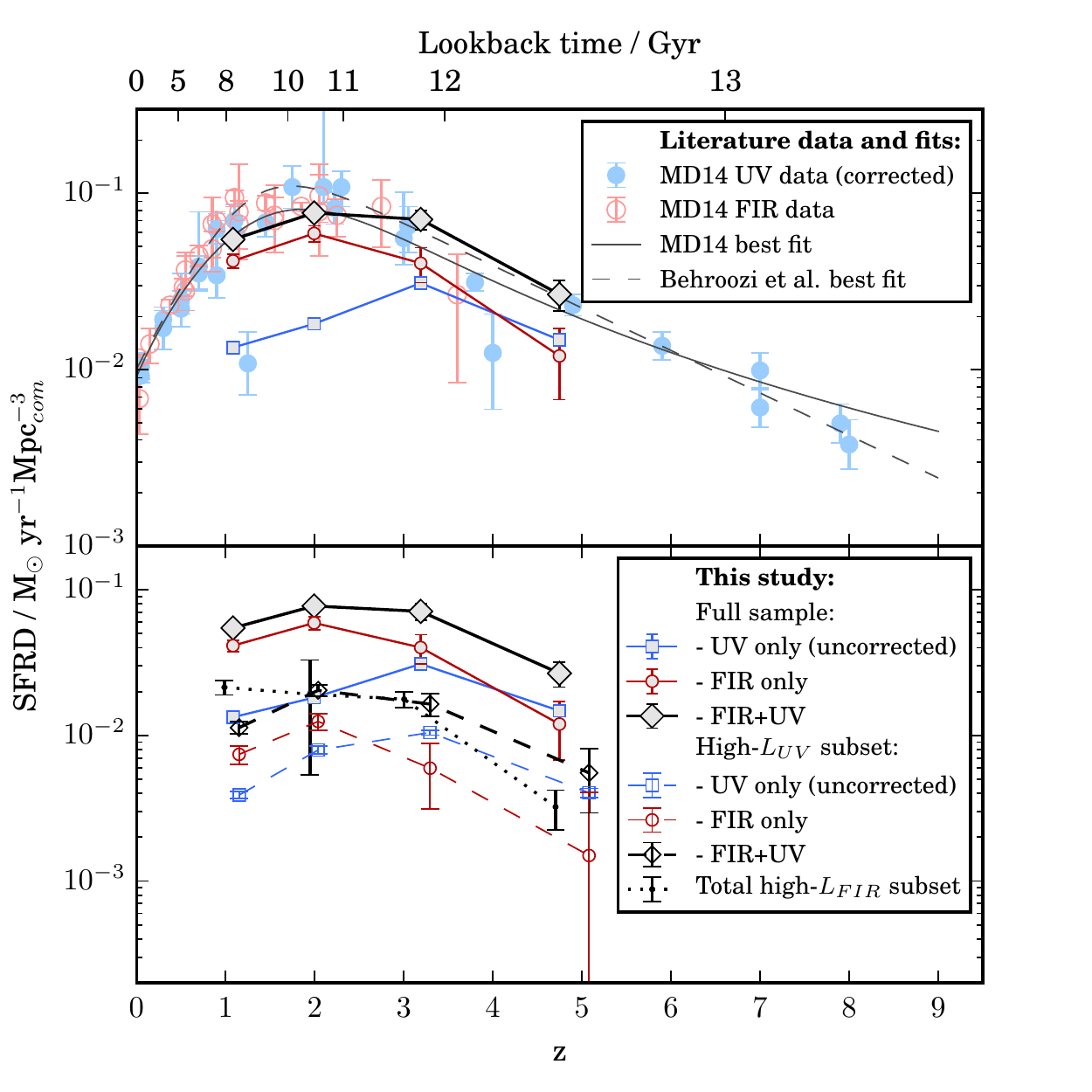}
\caption{Estimated total SFRD as a function of redshift, as in Fig.~\ref{fig:z_sfrd}, with corrections to account for the contribution from galaxies beneath the stellar-mass cut.
Both panels show the total SFRD (black solid lines), the unobscured portion from the rest-frame UV data (blue), and the obscured portion from the FIR (red). 
The lower panel also shows the contribution from the UV-luminosity-limited sample ($\MUV<M^{\ast}$; black dashed lines), broken down into unobscured (blue dashed) and obscured (red dashed); and the contribution from 450\micron-detected sources (black dotted line), which are almost completely obscured.
}
\label{fig:z_sfrd_lmc}
\end{center}
\end{figure*}

In Fig.~\ref{fig:z_sfrd} we plotted our direct measurements of the SFRD using a complete sample of massive galaxies with \mbox{$\Mstar>10^{10}\Msun$}. We now attempt to recover the full SFRD by estimating the additional contribution from galaxies with lower masses. The full unobscured SFRD up to high redshifts is well-characterised in the literature without explicit mass limits, by integration of the UV LF \citep{Bouwens2007,Bouwens2012,Bouwens2015, Cucciati2012, Smit2012, McLure2013, Duncan2014, Finkelstein2015, Mashian2015, Parsa2015, McLeod2016}. 
We use the UV LFs from \citet{Parsa2015} in CANDELS GOODS-S and the \emph{Hubble} Ultra-Deep Field (HUDF) to correct for the fraction of the unobscured SFRD that is missing from our direct measurements in Fig.~\ref{fig:z_sfrd}. 
To estimate this, we match the catalogue of UV absolute magnitudes from \citet{Parsa2015} with the catalogue of stellar masses in the same field from Mortlock et al. (2017, in preparation). 
The catalogue from \citeauthor{Parsa2015} contains photometric redshifts in the range $0<z<6$ (although only $1.5<z<4.5$ are included in their analysis).
In each of our redshift bins, we then measure (i) the integrated luminosity density of galaxies with $\Mstar>10^{10}\Msun$, down to the same \MUV\ limit as our bins (see Section~\ref{sec:binning}); and compare this with (ii) the integrated luminosity density in the full UV LF from \citet{Parsa2015}, down to a constant limit of $\MUV=-15$ at all redshifts. The ratio (ii)/(i) gives the correction factor to our direct measurement of the unobscured SFRD in our data set. 
Assuming a constant limiting $\MUV=-15$ at all redshifts facilitates comparisons across all redshifts. This limit was chosen because \citet{Parsa2015} showed that it effectively captures the full luminosity density.
The derived correction factors are 5.6, 8.0, 10.8 and 3.2 in the four redshift bins ($0.5<z<1.5$, $1.5<z<2.5$, $2.5<z<4$ and $4<z<6$) respectively.
We apply an analogous technique to correct our measurements of the unobscured SFRD in the UV-luminous ($\MUV<\MUV^\ast$) bins. The ratio of the total UV density at $\MUV<\MUV^\ast$ to that in massive galaxies ($\Mstar>10^{10}\Msun$) provides the correction factors of 3.3, 5.2, 7.5, 5.3 in the four redshift bins respectively.

The results, in Fig.~\ref{fig:z_sfrd_lmc}, show that accounting for galaxies with $\Mstar<10^{10}\Msun$ brings the unobscured SFRD in line with the obscured SFRD at $z>3$. This is because most of the unobscured star formation in the Universe occurs within low-mass galaxies, while the star formation in more massive galaxies is mostly obscured.
Now we see that while the unobscured SFRD peaks at $z\sim3$, the obscured (and therefore the total) SFRD peaks later, at $z\sim2$. While the obscured SFRD dominates at $z<3$, the obscured fraction appears to be close to 50 per cent at higher redshifts, and in UV-luminous galaxies the majority of the star formation is unobscured at $z>3$.
{A similar conclusion was obtained by \citet{Burgarella2013a} by analysing UV and IR LFs up to $z=4$.}

However, we have not yet accounted for the obscured SFRD at lower masses. The monotonic trends in Figs.~\ref{fig:mstar_sfr} and \ref{fig:irx_muv} (notwithstanding incompleteness-related bias at low masses) indicate that the average obscuration fraction (SFR$_\text{IR}$/SFR$_\text{UV}$) at $\Mstar<10^{10}\Msun$ is $\lesssim 1$ at all redshifts, and therefore that the SFRD at low masses is predominantly unobscured. While our sample is incomplete at low masses and high redshifts, we can nevertheless test this assertion by including all galaxies down to $10^{9}\Msun$ in our SFRD measurements, and calculating obscured SFRD from the sum of SFR$_\text{IR}/V_\text{max}$ (where $V_\text{max}$ is defined as the maximum volume within which that galaxy could reside and have been selected in our sample). Doing so, our obscured SFR values increase by a factor of roughly 1.25 in each redshift bin, a very small change to the data shown in Fig.~\ref{fig:z_sfrd_lmc} (note the logarithmic scale). By extension, the contribution to obscured SFRD from galaxies with $\Mstar<10^{9}\Msun$ can be assumed to be negligible.

Fig.~\ref{fig:z_sfrd_lmc} reveals a transition from an early Universe dominated by unobscured star formation to a Universe dominated by obscured star formation at ``cosmic noon'' (the peak of SFRD at $z\sim2.5$); yet Fig.~\ref{fig:mstar_sfr} indicates that obscuration at a given stellar mass is not redshift-dependent. 
The reason for this is the stellar-mass dependence of obscuration. At $z>3$ a greater fraction of the SFRD is contributed by lower-mass galaxies, which are less obscured, but by $z\sim2$ the high-mass end of the stellar mass function has built up and the SFRD is dominated by galaxies with $\Mstar>10^{10}\Msun$, which have heavily obscured star formation.

Several interesting comparisons can be made between our results in Fig.~\ref{fig:z_sfrd_lmc} and data from the literature. At $z<3$, our measurement of the total SFRD is very close to the fit to the literature compilation by \citet{Madau2014}, although at $z>3$ we measure slightly higher values that are closer to the functional form of \citet{Behroozi2013b}, and significantly lower than the earlier compilation of \citep[not shown here]{Hopkins2006}.\footnote{Note that the offset between the \citet{Behroozi2013b} curve and the \citet{Madau2014} curve may be explained by the fact that the former is estimated from the integral over all halo masses in their model, while the latter is estimated by integrating the IR and UV LFs down to $0.03\,L^{\ast}$.} 
A great deal of scatter is seen between different studies around the peak of the cosmic star formation history \citep[for a review, see][]{Behroozi2013b,Madau2014}, as a result of the broad range of observational techniques that have been employed (IR+UV; dust-corrected UV; 1.4GHz; H$\alpha$; etc). In contrast, the chief tracer that has been available at $z>4$ is the UV LF, which relies on an assumed universal IRX--$\beta$ relationship. In Section~\ref{sec:irxbeta} we showed that this relationship may be mass-dependent and is not necessarily always consistent with the \citet{Meurer1999} relation that is commonly assumed \citep[an issue raised by many other high-redshift studies, e.g.][]{Buat2012,Casey2014,Oteo2014,Pannella2015,Talia2015}.
In spite of this, and in spite of the significant deviations between the dust-corrected UV SSFRs and the UV+IR SSFRs in Fig.~\ref{fig:z_ssfr}, we find that the total UV+IR SFRD in Fig.~\ref{fig:z_sfrd_lmc} is surprisingly close to the UV-based determinations at $z>4$ from studies such as \citet{Bouwens2012}, which rely on an assumed dust correction based on \citet{Meurer1999}. In fact, the increasing dominance of the unobscured SFRD between $z\sim3$ and $z\sim5$ can only strengthen our confidence in determining the continuing evolution of the SFRD towards $z\sim10$. 

On the other hand, accounting for obscuration is key to accurately measuring the SFRD at $1<z<4$.
In comparison to our results, \citet{Burgarella2013a} measured a higher obscured SFRD (and a higher obscured/unobscured ratio) at $2<z<4$, by integrating FIR LFs from \Herschel\ \citep{Gruppioni2013} and UV LFs from \citet{Cucciati2012}. Their results are particularly subject to uncertainties in the faint end of both LFs at $z>2$, and are consistent with our results within these uncertainties.
\citet{Viero2013} and \citet{Schreiber2015} both measured the obscured SFRD in stacked \Herschel\ data at $0<z<4$. 
\citeauthor{Viero2013} found that it closely followed the \citet{Behroozi2013b} model, which is slightly above our estimates for the obscured SFRD (but is consistent with our estimate of the \emph{total} SFRD).
This slight discrepancy may result from systematic uncertainties in the FIR SEDs, since \citeauthor{Viero2013} fitted their data with SEDs that had temperatures increasing with redshift, while we assumed a non-evolving SED based on our best-fitting templates (our error bars account for this uncertainty; {see Section~\ref{sec:sfrcalib}}).
\citet{Schreiber2015} similarly measured the total SFRD by stacking \Herschel\ data at $0<z<4$, and adding in the unobscured UV contribution. Their results closely followed the fits from \citet{Madau2014}, and are consistent with our own.
Our results also agree with the measurements of the obscured+unobscured SFRD at $1<z<5$ from stacking in the ALMA map of the HUDF \citep{Dunlop2016}.
Furthermore, independent measurements of the total SFRD at $2<z<5$ from O\textsc{ii} emission-line surveys \citep{Khostovan2015} are consistent with our determinations (once accounting for the different IMF assumed in that study).

\section{Conclusions}
In this paper we have demonstrated how statistical information about faint, high-redshift source populations can be extracted from confused, single-dish, submm surveys (from S2CLS) with a combination of deep, value-added positional prior catalogues (from CANDELS/3D-HST) and the computational deconfusion technique offered by \tphot\ ({described in} Sections~\ref{sec:methods}--\ref{sec:analysis}).
{We applied these techniques to 230~arcmin$^2$ of the deepest 450-\micron\ imaging over the AEGIS, COSMOS and UDS CANDELS fields, in order to measure the obscured SFRs of stellar-mass-selected galaxies at $0.5<z<6$. We used additional data at 100--850\micron\ to constrain the FIR SED, which we modelled with a single average template at all redshifts. We cannot exclude the possibility of evolution in the SED shape, but due to our use of submm data close to the SED peak at all redshifts, the resulting systematic uncertainties are small and are fully accounted for in our errors (see Sections~\ref{sec:sedfit}, \ref{sec:sfrcalib}).
We obtained the following main results:
}
\begin{enumerate}
\item We detect 165 galaxies at 450\micron\ with $S/N>3$ at $S_{450}\gtrsim 3$~mJy, similar to published 450-\micron\ samples from SCUBA-2. The detected sources have a broad redshift distribution at $0<z<4$ (median $z=1.68$), although we also detect four sources at $4<z<6$. They span a wide range of stellar masses (typically $9.5<\log(\Mstar/\Msun)<11.5$) and UV luminosities (typically $-21<\MUV<-16$). This sample generally traces the highest SFRs at $z<4$, but {exhibits} a wide range of obscuration fractions, with $1<\log(\Lir/\Luv)<4$ (see Section~\ref{sec:sfr}).
\item In the stacked results, total SFR (from IR+UV {data}) is strongly correlated with stellar mass at all redshifts, while the raw UV luminosity is a relatively weak indicator of total SFR, especially at $z\gtrsim2$ (Section~\ref{sec:sfr}).
\item Instead, UV luminosity primarily indicates the level of SFR obscuration {at a given stellar mass}. 
{The mean obscuration is strongly correlated with both stellar mass and UV luminosity, but does not appear to evolve significantly with redshift at a given \Mstar\ and \Luv, at least in our stellar-mass-selected sample (Sections~\ref{sec:sfr}, \ref{sec:IRXMstarMuv}).}
{When restricting the analysis to $UVJ$-selected star-forming galaxies, we} 
find that the obscuration can be determined from the stellar mass and UV luminosity as \mbox{$\log(\Lir/\Luv) = a+b(\Luv/10^9\Lsun)+c(\Mstar/10^9\Msun)$}, where \mbox{$a=5.9\pm1.8$}, \mbox{$b=-0.56\pm0.07$}, \mbox{$c=0.70\pm0.07$}.
\item The average UV+IR SSFRs of $UVJ$-selected star-forming galaxies rise with redshift, and the evolution is steeper for more massive galaxies, indicating that, on average, they stop forming stars earlier.
Massive galaxies ($\Mstar>10^{10}\Msun$) have average SSFRs $\sim1-2$~Gyr$^{-1}$ at $z\sim2-4$, and $\sim3-4$~Gyr$^{-1}$ at $z\sim5$, in agreement with the most recent studies from both SCUBA-2 and ALMA.
We fit the binned data with a bivariate model: 
$\log(\text{SSFR}/\text{Gyr}^{-1})=a(z)[\log(\Mstar/\Msun)-10.5]+b(z)$.
The evolution of the slope of the SSFR($\Mstar$) {relation} is given by 
$a(z)=(-0.64\pm0.19) + (0.76\pm0.45)\log(1+z)$, while the evolution of the normalization (at $\log(\Mstar/\Msun)=10.5$) is given by 
$b(z)=(-9.57\pm0.11) + (1.59\pm0.24)\log(1+z)$
(Section~\ref{sec:ssfr}). 
\item Dust-corrected SFRs from the UV luminosity and spectral slope ($\beta$) can under-estimate the total SFR and overall predict a weaker evolution in the average SSFR as a function of redshift (Section~\ref{sec:uvcorr}). 
\item By stacking $UVJ$-selected star-forming galaxies, we find that massive galaxies ($\Mstar>10^{11}\Msun$) tend to have higher obscuration for a given $\beta$ in comparison with lower-mass galaxies. It is also possible that the IRX--$\beta$ relation evolves with redshift, although this result could be influenced by contamination of our $z<1.5$ bins with passive galaxies (Section~\ref{sec:irxbeta}).
\item Our results provide homogeneous measurements of the obscured SFR density (SFRD) in a highly-complete sample of massive galaxies ($\Mstar>10^{10}\Msun$) over the redshift range $0.5<z<6$, extending beyond what has been possible in previous studies using \Herschel\ data. 
We show that obscured star formation dominates the total SFRD in massive galaxies at all redshifts, and exceeds unobscured star formation by a factor $>10$ (Section~\ref{sec:sfrd}). 
\item The FIR-detected sample, which is effectively flux-limited at $S_{450}>3$~mJy and samples the highest star-formation rates at all redshifts (SFR~$\gtrsim100\sfr$), accounts for approximately one third of the total SFRD over the redshift range $0.5<z<6$.
\item The most UV-luminous massive galaxies, defined as those with $\MUV<\MUV^{\ast}$ and $\Mstar>10^{10}\Msun$, account for around one fifth of the total SFRD over the same redshift range, but even in these the majority of the SFRD is obscured at $z\lesssim3$.
\item After correcting for the contributions from lower-mass galaxies, the full SFRD from UV+IR data is in good agreement with previous literature estimates both from UV+IR at $z\lesssim3$ and from UV-only data at $z\sim5$. This indicates that UV-selected samples with $\beta$ dust corrections are successful for calibrating total SFRD at the highest redshifts ($z>3$), in spite of variations in the IRX--$\beta$ relation and the lower SSFRs estimated from the dust-corrected UV alone. This is due to the increasing dominance of unobscured star formation at $z\gtrsim3$ (Section~\ref{sec:lmc}). 
\item When accounting for all stellar masses, the SFRD at $z\lesssim3$ remains dominated by obscured star formation, but at higher redshifts the total obscured and unobscured SFRD are equal, while the most UV-luminous galaxies are predominantly unobscured at $z\gtrsim3$. The SFRD contribution from the most UV-luminous and the FIR-detected galaxies are approximately equal at $z\sim$~2--3 when including all stellar masses.
\item We conclude that the reason for this transition from an early Universe ($z>3$) dominated by unobscured star formation to a Universe dominated by obscured star formation at cosmic noon ($z\approx2$) is explained by the increasing contribution of massive galaxies as the high-mass end of the stellar mass function is built up at around $z\sim$~2--3. This is consistent with our observation that the obscuration of star formation at a fixed stellar mass is independent of redshift.
\end{enumerate}

\section*{Acknowledgements}
{The authors would like to warmly thank the referee, V. Buat, for providing a careful and constructive report which significantly improved the paper.}
The research leading to these results has received funding from the European Union Seventh Framework Programme (FP7/2007-2013) under grant agreement no. 312725, and also from the European Research Council via the award of an Advanced Grant (JSD).
KC acknowledges support from STFC (grant number ST/M001008/1).
KK acknowledges support from the Swedish Research Council.
The sub-millimetre data used in this paper were obtained from the JCMT SCUBA-2 Cosmology Legacy Survey.
The James Clerk Maxwell Telescope has historically been operated by the Joint Astronomy Centre on behalf of the Science and Technology Facilities Council of the United Kingdom, the National Research Council of Canada and the Netherlands Organisation for Scientific Research. 
Additional funds for the construction of SCUBA-2 were provided by the Canada Foundation for Innovation. 
This research has also made use of data from HerMES project (\url{http://hermes.sussex.ac.uk/}). HerMES is a Herschel Key Programme utilising Guaranteed Time from the SPIRE instrument team, ESAC scientists and a mission scientist.
The HerMES data was accessed through the Herschel Database in Marseille (HeDaM - \url{http://hedam.lam.fr}) operated by CeSAM and hosted by the Laboratoire d'Astrophysique de Marseille.
This work uses observations taken by the 3D-HST Treasury Program (GO 12177 and 12328) with the NASA/ESA HST, which is operated by the Association of Universities for Research in Astronomy, Inc., under NASA contract NAS5-26555.
The analysis in this paper has made use of the open-source Python packages \textsc{matplotlib} \citep{Hunter2007} and \textsc{astropy}, a community-developed core Python package for Astronomy \citep{astropy2013}.


\bibliographystyle{mn2e}
\bibliography{MyLibrary20120814}



\appendix
\section{Stacked SFRs of star-forming galaxies}
\label{app:uvj}
\begin{figure*}
\begin{center}
\includegraphics[width=\textwidth,clip,trim=2cm 35mm 5mm 1cm]{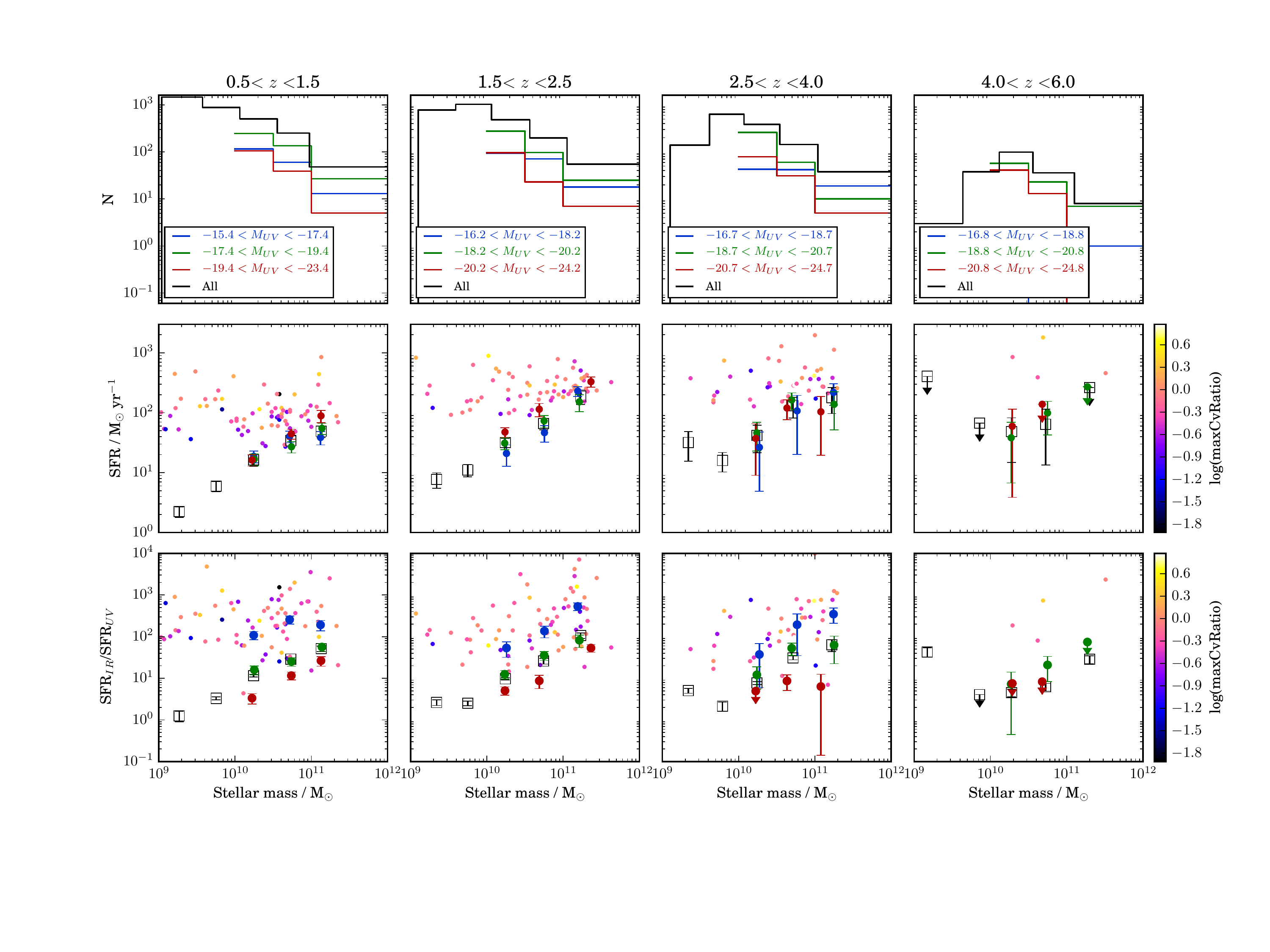}
\caption{Same as Fig.~\ref{fig:mstar_sfr}, but excluding passive galaxies as defined by \eqnref{eqn:uvj}. 
From top to bottom:
Number $N$ per bin; total SFR$_\text{UV}$~+~SFR$_\text{IR}$; and obscuration ratio SFR$_\text{IR}$/SFR$_\text{UV}$. 
Large black squares show the full mass-binned stacks, while large filled symbols with error bars show the stacks divided into bins of $\MUV$, defined relative to $\MUV^\ast$ at the appropriate redshift \citep{Parsa2015}.
FIR detections ($S/N>3$) are shown by small coloured points in which the colour coding indicates \emph{maxCvRatio}.
}
\label{fig:mstar_sfr_uvj}
\end{center}
\end{figure*}

\begin{figure*}
\begin{center}
\includegraphics[width=\textwidth,clip,trim=2cm 5mm 5mm 0mm]{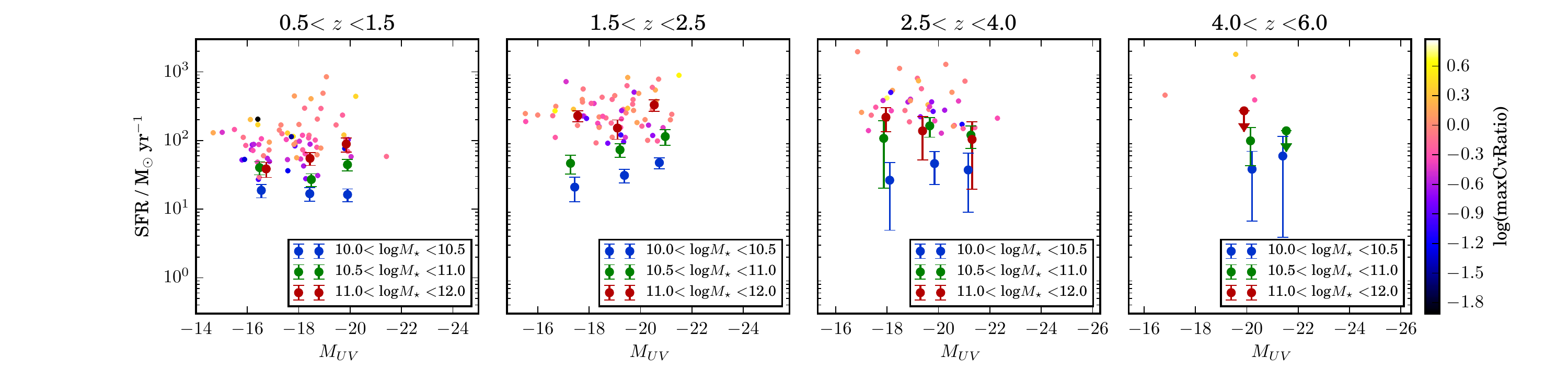}
\caption{Same as Fig.~\ref{fig:muv_sfr}, but excluding passive galaxies as defined by \eqnref{eqn:uvj}. 
The data plotted are the same as in Fig~\ref{fig:mstar_sfr_uvj}, except that colours here indicate the stellar-mass bin. 
}
\label{fig:muv_sfr_uvj}
\end{center}
\end{figure*}

In Section~\ref{sec:sfr} we showed the results of stacked SFRs from IR+UV data encompassing all galaxies in the sample, divided into bins of redshift, stellar mass and UV absolute magnitude. However, in Section~\ref{sec:ssfr} we used $UVJ$ colour criteria to remove passive galaxies from the sample in order to investigate the SSFR and IR/UV luminosity ratio of star-forming galaxies as a function of stellar mass and redshift. 
To show the effects of removing passive galaxies on the average SFR as a function of redshift, stellar mass and \MUV, we reproduce Figs.~\ref{fig:mstar_sfr} and \ref{fig:muv_sfr} after applying the $UVJ$ cuts in \eqnref{eqn:uvj} to remove passive galaxies. The results are shown as a function of stellar mass in Fig.~\ref{fig:mstar_sfr_uvj}, and as a function of \MUV\ in Fig.~\ref{fig:muv_sfr_uvj}.

\section{Example T-PHOT parameter file}
\label{app:tphotparfile}
\begin{verbatim}
# TPHOT PARAMETER FILE

#------------------- PIPELINE --------------------#
# Select the desired one or create another:
# 1st PASS:
order           positions, fit, diags, archive

#----------------- CUTOUT STAGE ------------------#
poscat          3DHST_xy_positions.dat
culling         False
relscale        1
cutoutdir       cutouts
cutoutcat       cutouts/_cutouts.cat
normalize       true

#----------------- CONVOLUTION STAGE -------------#

loresfile       AEGIS-CANDELS-4_flux.fits
loreserr        AEGIS-CANDELS-4_rms.fits
errtype         rms
rmsconstant     1
bgconstant      0
maxflag         64            
FFTconv		        true
multikernels	   false
kernelfile      psf-4.fits
psffile      	  psf-4.fits
kernellookup    ch1_dancecard.txt
posframe        hires
templatedir     templates
templatecat	    templates/_templates.cat

#----------------- FITTING STAGE ------------------#

# Filenames:
fitpars	         tpipe_tphot.param
tphotcat         lores_tphot.cat_pass1
tphotcell        lores_tphot.cell_pass1
tphotcovar       lores_tphot.covar_pass1
# Control parameters:
fitting 	        single
dithercell      true
cellmask        true
maskfloor       1e-9
fitbackground   true
writecovar      true
threshold	0.0
linsyssolver	   lu # [options: ibg, cholesky, lu] 
clip		           false

#------------------ DIAGNOSTICS STAGES ------------#

modelfile       lores_collage_pass1.fits
# Dance:
dzonesize       100
maxshift        1.0
ddiagfile       ddiags.txt
dlogfile        dlog.txt
dancefft	       true
\end{verbatim}

\bsp	
\label{lastpage}
\end{document}